\documentclass{emulateapj}
\usepackage{natbib}
\usepackage{rotating}
\usepackage{mathtools}
\usepackage{color} 

\begin{document}

\title{Neutrino-driven winds in the aftermath of a neutron star merger: nucleosynthesis and electromagnetic transients}


\author{D. Martin}
\author{A. Perego}
\author{A. Arcones}
\affil{Institut f\"ur Kernphysik, Technische Universit\"at Darmstadt, Schlossgartenstr. 2,
Darmstadt D-64289, Germany\\
\and GSI Helmholtzzentrum f\"ur Schwerionenforschung GmbH, Planckstr. 1, Darmstadt D-64291, Germany}
\email{dirk.martin@physik.tu-darmstadt.de}
\author{F.-K. Thielemann} 
\affil{Department of Physics, University of Basel, Klingelbergstra{\ss}e 82, 4056, Basel, Switzerland}
\author{O. Korobkin}
\author{S. Rosswog} 
\affil{The Oskar Klein Centre, Department of Astronomy, AlbaNova, Stockholm University, SE-106 91 Stockholm, Sweden}

\date{today}

\begin{abstract}
We present a comprehensive nucleosynthesis study of the neutrino-driven wind in the
aftermath of a binary neutron star merger. Our focus is the initial remnant phase when 
a massive central neutron star is present. Using tracers from a recent hydrodynamical simulation, we
determine total masses and integrated abundances to characterize the
composition of unbound matter. We find that the nucleosynthetic yields depend sensitively on both the
life time of the massive neutron star and the polar angle. Matter in excess
of up to $9 \cdot 10^{-3} M_\odot$ becomes unbound until $\sim 200$~ms. Due to electron
fractions of $Y_{\rm e} \approx 0.2 - 0.4$ mainly nuclei with mass numbers $A < 130$
are synthesized, complementing the yields from the earlier dynamic ejecta. Mixing
scenarios with these two types of ejecta can explain the abundance pattern in r-process
enriched metal-poor stars. Additionally, we calculate heating rates for the decay of 
the freshly produced radioactive isotopes. The resulting light curve peaks in the blue 
band after about 4~h. Furthermore, high opacities due to heavy r-process nuclei in the 
dynamic ejecta lead to a second peak in the infrared after $3-4$~d.
\end{abstract}

\keywords{accretion, accretion disks --- nuclear reactions, nucleosynthesis, abundances --- neutrinos --- stars: neutron}
\section{Introduction}
\label{sec:intro}

Since the first detection of neutron star binary systems, the
merger of two neutron stars or a neutron star with a black hole
has been recognized as an ideal environment for 
rapid neutron capture (r-process) nucleosynthesis 
\citep{Lattimer.Schramm:1974,Lattimer.Schramm:1976,Eichler.etal:1989}.
In the r-process, the neutron density is so high that neutrons are 
captured much faster than the involved nuclei can decay. Once the neutron supply becomes 
insufficient, the temporarily formed, extremely neutron-rich nuclei decay
toward the valley of beta-stability. Approximately
half of all the elements heavier than iron are produced in this way. 

Matter that is ejected by hydrodynamical interaction and gravitational torques,
often referred to as dynamic ejecta, has long been recognized as the likely
source of the {\em heaviest} r-process nuclei \citep{Freiburghaus.etal:1999a,Rosswog.etal:1999,Roberts.etal:2011,Goriely.etal:2011,
Korobkin.etal:2012,Bauswein.etal:2013,Hotokezaka.etal:2013,Rosswog:2013,Rosswog.etal:2014}. 
For given nuclear physics input, this type of ejecta yields a very robust, 
``strong r-process'' \citep{Sneden.etal:2008} pattern due to efficient fission cycling \citep{Panov.etal:2008,Goriely.etal:2011,Korobkin.etal:2012,Petermann.etal:2012,
Bauswein.etal:2013,Wanajo.etal:2014,Goriely:2015,Just.etal:2015a}.
This means that all merger events produce practically identical abundance patterns (for this
type of ejecta), independent of the details of the merging system. The abundance pattern,
however, shows some sensitivity to the resulting distribution of fission fragments
and to the resulting beta-decay half lives \citep{Goriely.etal:2013,Eichler.etal:2015,Goriely:2015}.

With increasing sophistication of the included physics and of the
numerical techniques also our view on the merger process and its mass
loss channels has sharpened.  For example, the dynamic ejecta have
been recognized to consist of a cold ``tidal component'' that is
launched via gravitational torques and a hotter ``interaction
component'' that comes from the hydrodynamical interaction of the
neutron stars \citep{Oechslin.etal:2007,
  Korobkin.etal:2012}. 
Recent general relativity (GR) simulations of the dynamic ejecta
including neutrino transport \citep{Sekiguchi.etal:2015,Foucart.etal:2015}
show that shocks in the interaction component significantly increase the
temperature. This hotter environment enhances the weak reaction rates.
In particular, the larger positron capture and electron neutrino absorption rates
on neutrons increase the electron fraction, potentially up to
values of 0.4. Therefore, the nucleosynthesis based on these general 
relativistic simulations can produce the full
r-process from the first peak ($A=80$) to the third one ($A=195)$,
see also \cite{Goriely.etal:2015}. In contrast, Newtonian
simulations or approximate GR simulations without weak reactions
\cite{Goriely.etal:2011,Bauswein.etal:2013} only produce
the heavy r-process elements ($A \geq 130$).
  
In addition to the dynamic ejecta, the presence of two more mass loss
channels has been appreciated and begun to be explored in more
detail. These are neutrino-driven winds
\citep{Ruffert.etal:1997,Rosswog.Ramirez-Ruiz:2002, Dessart.etal:2009,
  Perego.etal:2014, Just.etal:2015a} and late-time accretion disk	
disintegration \citep{Metzger.etal:2008, Beloborodov:2008,
  Lee.etal:2009, Fernandez.Metzger:2013, Just.etal:2015a}. 
The late-time disintegration of the accretion disk, driven by 
viscous heating and recombination of free nucleons into alpha particles, 
is referred to as viscous ejecta in the following.

The historically favored production site, core-collapse supernovae, in
contrast, seems to struggle to produce the conditions required for a
successful r-process \citep{Arcones.etal:2007, Huedepohl.etal:2010,
  Roberts.etal:2010, Fischer.etal:2010, Arcones.Montes:2011,
  Arcones.Thielemann:2013}, though some rare supernovae may plausibly
contribute \citep{Fujimoto.etal:2006, Fujimoto.etal:2008,
  Winteler.etal:2012, Moesta.etal:2014, Nishimura.etal:2015}.

The idea that compact binary mergers could be the sources of the
heaviest r-process elements has recently received strong support from
an observed near-infrared transient in the aftermath of a short
gamma-ray burst (sGRB) \citep{Tanvir.etal:2013, Berger.etal:2013,
  Yang.etal:2015}.  This emission has been interpreted as being a
``kilonova'' \citep{Metzger.etal:2010b} (sometimes referred to as
``macronova'', cf. \cite{Kulkarni:2005}), a transient that is powered
by the radioactive decay of freshly synthesized r-process elements. In
particular the luminosity peak after only several days and in the
near-infrared -- rather than after hours in the optical/UV as
originally expected \citep{Li.Paczynski:1998} -- supports the view
that the ejecta are made of very heavy r-process elements that result
in effective opacities that are orders of magnitude larger than those
of supernovae \citep{Kasen.etal:2013, Barnes.Kasen:2013,
  Tanaka.Hotokezaka:2013, Grossman.etal:2014}.  If this interpretation
is correct, this would for the first time link compact binary mergers
with GRBs observationally, a long suspected, but actually so far
unproven connection, and also with the formation of very heavy
elements. Such kilonovae are an important facet in the multi-messenger
view of compact binary mergers \citep{Rosswog:2015} and, since they
emit quasi-isotropically, they may become crucial in assuring the
first direct gravitational wave detections
\citep{Metzger.Berger:2012,Nissanke.etal:2013,Piran.etal:2013}.

Despite all of these promises, the role of compact binary mergers for 
the chemical enrichment histories of galaxies, in particular at early times, is not
yet sufficiently well understood. While earlier work \citep{Argast.etal:2004} disfavored 
them as dominant source of r-process in the early galaxy, more recent studies are either more 
optimistic about their role \citep{Matteucci.etal:2014} or actually favor them 
over supernovae as major production site, see for example the  recent hydrodynamical studies 
of \cite{Shen.etal:2015} and \cite{Vandevoort.etal:2014}. Here more work is clearly indicated
to settle the case (see \cite{Cescutti.etal:2015,Wehmeyer.etal:2015}).

Understanding the different mass loss channels, their nucleosynthesis
and possible electromagnetic transients is of prime importance for
cosmic nucleosynthesis and galactical chemical evolution, for
gravitational wave detection strategies and actually also for the GRB
launch mechanism, since already a tiny mass loading of a fireball
can ``choke'' an emerging jet \citep{Murguia.etal:2014}. In this
paper, we focus on the neutrino-driven wind from a post-merger remnant
that consists of a central massive neutron star (MNS) surrounded by an
accretion disk and present a comprehensive nucleosynthesis study based on the first three
dimensional simulation recently performed by \cite{Perego.etal:2014}.
In the latter work preliminary nucleosynthesis results based on ten
representative trajectories were presented. Here, we follow the
nucleosynthetic evolution based on $17\,000$ trajectories and we discuss
the implications for radioactively powered electromagnetic transients.

The paper is structured in the following way. In Sect.~\ref{sec:methods} 
we briefly describe the hydrodynamical simulations, the tracers used to 
follow the ejecta and the physical input of the reaction network. We
present the results of the nucleosynthesis calculations and report on 
time and angle dependency of the neutrino-driven wind in Sect.~\ref{sec:results}. 
Furthermore, we consider mixing with the other channels of matter 
ejection and calculate the electromagnetic signal. Finally, we discuss 
our results and their implications in Sect.~\ref{sec:conclusions}.

\section{Methods}
\label{sec:methods}

\subsection{Hydrodynamical simulations}
\label{subsec:hydro}
Our nucleosynthesis study is based on the first three dimensional,
Newtonian simulation of the neutrino-driven wind that emerges during
the aftermath of a binary neutron star merger
\citep{Perego.etal:2014}. The merger remnant is characterized by a
long-lived massive neutron star, surrounded by a quasi-Keplerian
accretion disk with an inital mass of $M_{\rm disk} \simeq 0.19 M_\odot$. 
For the present analysis, we repeat and extend the
previous simulation to longer times, and we follow a substantially
larger number of particles.

We use the parallel grid code \texttt{FISH} \citep{Kaeppeli.etal:2011} to solve
the hydrodynamical equations on a uniform Cartesian grid. Nuclear
matter description is provided by the TM1 nuclear equation of state,
supplemented with electron-positron and photon contributions \citep{Timmes.Swesty:2000,Hempel.etal:2012}.
As initial condition we use a late matter configuration from a three
dimensional high resolution simulation of two non-spinning 1.4 ${\rm M}_{\odot}$ 
neutron stars (e.g. \cite{Price.Rosswog:2006}).
Neutrino-matter interactions are taken into account using the
multiflavor Advanced Spectral Leakage (ASL) scheme, see Perego et al. (2015, in preparation). It
models the neutrino emission effectively by smoothly interpolating between
diffusive and free-streaming rates, separately for different neutrino
energies, and it has been carefully gauged at full transport calculations. 
Neutrino absorption is also included in optically thin
conditions, based on the calculation of the neutrino densities outside
the neutrino last scattering surfaces. The list of the neutrino
reactions implemented inside the ASL scheme can be found in table~1 of
\cite{Perego.etal:2014}.
The present simulation corresponds to an extension of the one
  presented in \cite{Perego.etal:2014}. The simulations starts at 25 ms
after the beginning of the merger, including 10~ms of relaxation 
of the SPH final conditions on the grid, and follows the post merger 
evolution for 190~ms. The temporal evolution of the neutrino
  luminosities and mean energies during the whole simulation follows
  the same trends reported in Fig.~10 of \cite{Perego.etal:2014} for the
  first 90~ms. The relevant mean energies stay approximately constant:
  $\sim$11~MeV for the electron neutrinos and $\sim$15.5~MeV for the
  electron antineutrinos. The total neutrino luminosities, integrated
  over the whole solid angle, decrease slowly with time. At the end of
  the simulation ($t_\mathrm{sim} \approx 190$~ms) the electron (anti)neutrino
  luminosity due to cooling processes only is equal to $\sim 2.2 \cdot
  10^{52}$~erg/s ($\sim 2.9 \cdot 10^{52}$~erg/s), while the inclusion of the
  neutrino absorption in optically thin conditions decreases it to 
  $\sim 1.5 \cdot 10^{52}$~erg/s ($\sim 2.4 \cdot 10^{52}$~erg/s). 
  Due to the larger optical
  depth along the equatorial plane, the neutrino fluxes measured far
  from the MNS along the polar direction are roughly 3~times
  larger than the ones in the equatorial plane (cf. Fig.~12 of \cite{Perego.etal:2014}).

\begin{figure*}[!bht]
  \begin{center}
    \includegraphics[width=0.95\textwidth]{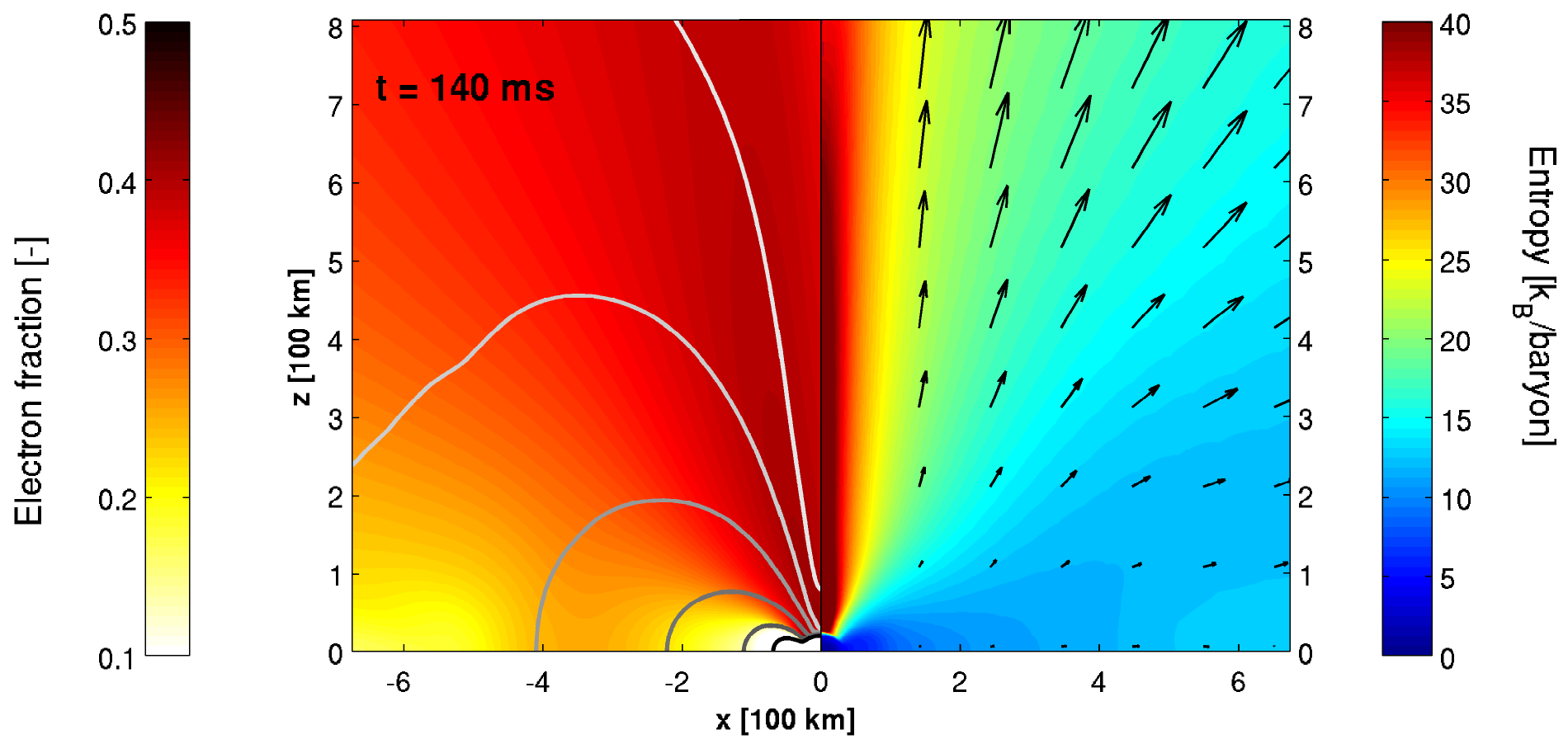}
  \end{center}
  \caption{$x-z$ plane at 140~ms after the beginning of the
    simulation. The left panel shows the electron fraction with a superimposed
    contour plot of the density. A color gradient from white to black indicates
    the regions with densities of $\rho = 10^6~{\rm g/cm}^3$ to $\rho = 10^{11}~{\rm g/cm}^3$. 
    Entropy profile and the projected velocity are presented in the right panel. 
    The length of the arrows characterizes the magnitude of the velocity.}
  \label{fig:simulation-rho-v}
\end{figure*}

The inclusion of the energy deposition provided by the neutrino
absorption on nucleons inside the disk drives a
baryonic wind on a time scale of tens of milliseconds, see
Fig.~\ref{fig:simulation-rho-v}. Matter expanding inside the wind
becomes unbound at a distance $\lesssim 600 \, {\rm km}$ from the
center. The resulting ejecta are confined within a polar angle of
$60^\circ$, measured from the rotational axis of the disk.
Due to the strong neutrino irradiation, the initially highly neutron
rich matter inside the disk changes its electron fraction in the
wind. The dominant $\nu_e$-absorption of neutrons raises $Y_e$ toward
equilibrium values (see e.g. \cite{Qian.Woosley:1996}). The
  evolution towards this equilibrium value may be affected by general
  relativistic treatment of the merger phase. 
  GR simulations of a binary neutron star merger including neutrinos \citep{Neilsen.etal:2014}
  and following the onset of the neutrino-driven wind 
  \citep{Sekiguchi.etal:2015} indicate that the luminosities can be 
  larger than in Newtonian simulations, due to the higher temperature 
  inside the massive neutron star and the disk.
  However, the ratio between the electron neutrino and antineutrino 
  luminosities is similar in both cases, as well as the values of the 
  mean energies. Therefore, the equilibrium electron fraction is 
  expected to be almost the same, while the evolution towards this 
  $Y_{e}$ should be faster in the GR simulations. This 
  difference does not come directly from the Newtonian treatment of 
  the disk and of the wind, but from the initial profiles resulting 
  from the merger process.

\subsection{Tracers from the hydrodynamical simulation}
\label{subsec:hydro-results}
\begin{figure}[!htb]
  \begin{center}
    \includegraphics[width=0.95\linewidth]{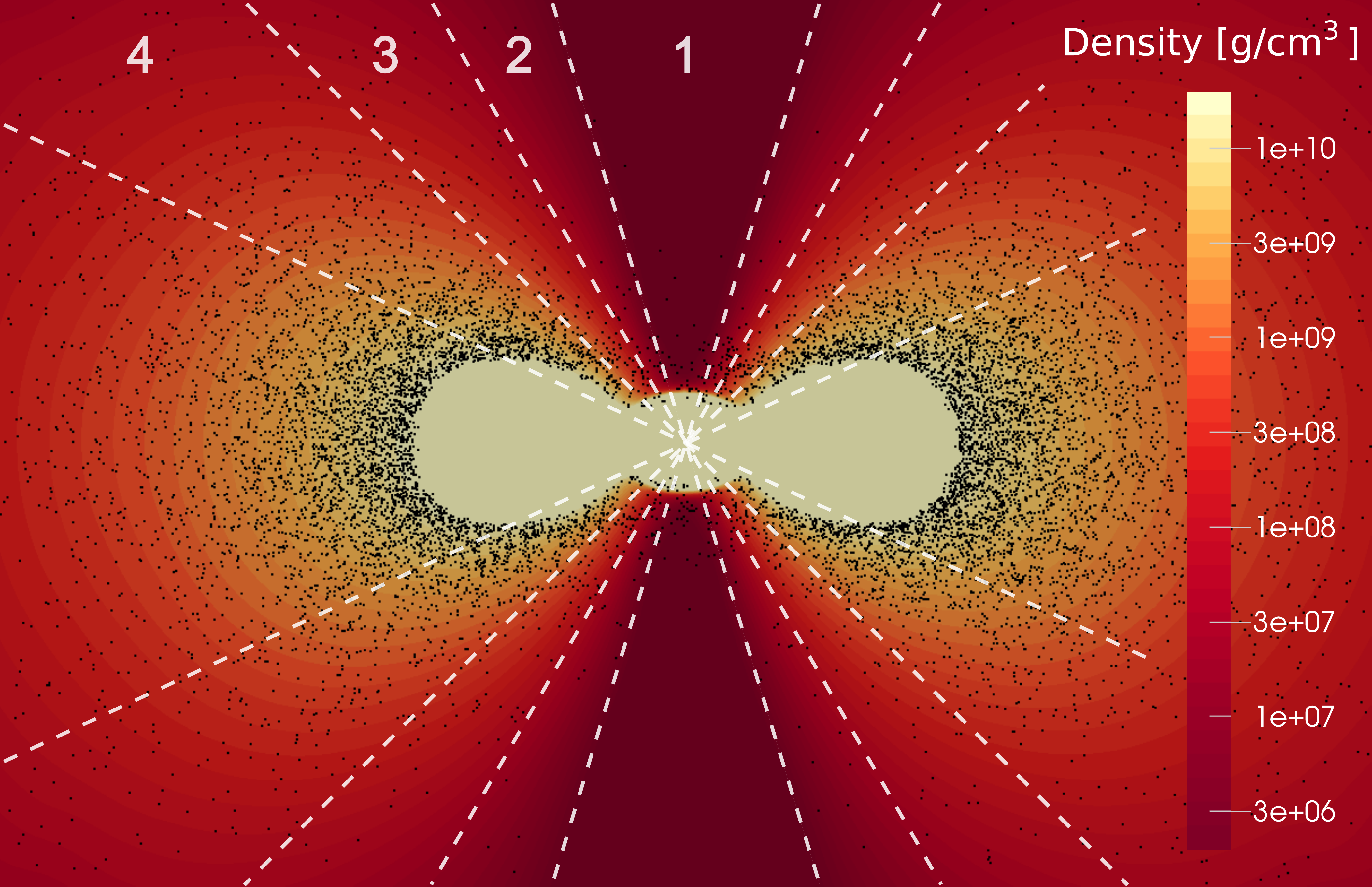}
  \end{center}
  \caption{Distribution of the tracers in the $x-z$ plane at the beginning of the
    simulation. The density profile is shown for \mbox{$2 \cdot 10^{6} \, {\rm g \, cm^{-3}} \leq
\rho \leq 2 \cdot 10^{10} \, {\rm g \, cm^{-3}}$}. Tracers are marked by black dots, 
    white dashed lines and numbers label the four angular regions of interest, which are referred 
    to as (angular) bins~$1-4$ in the following.}
  \label{fig:tracer-xz}
\end{figure}

At the beginning of our simulation we place $10^5$ tracers inside our
computational domain. The wind tracers are ejected from a region with
initial density of $\rho \lesssim 10^{10} \, {\rm g \, cm^{-3} }$, as
found in backtracking procedures of preliminary tests following
\cite{Perego.etal:2014}. Thus, we locate our tracers initially inside
the thick accretion disk where the density is between $ 2 \cdot 10^{6} \, {\rm g \,
  cm^{-3}}$ and $2 \cdot 10^{10} \, {\rm g \, cm^{-3}}$ (see
Fig.~\ref{fig:tracer-xz}). The number of tracers assigned to each grid
cell is proportional to its mass content and the actual location of
each particle within a cell is randomly assigned. Hence, the resulting tracer
distribution tracks the matter density distribution inside the disk.
The mass of the region of the disk where the tracers are placed is
0.0542 $M_\odot$. Therefore, we assign to each of them an inherent
mass of $m_i = m = 5.420 \cdot 10^{-7} M_\odot$. Every tracer records the local properties of
matter (density, internal energy, electron fraction and velocity
components) at its present location by linearly interpolating the
corresponding values on the computational grid. The total specific
energy, $e_{\rm tot}$, is computed as the algebraic sum of the
kinetic, thermal and gravitational specific energy. We recall here
that the thermal energy is obtained from the relativistic internal
energy, reduced by the rest mass energy (both obtained from the
nuclear equation of state). All tracers are passively
advected by the fluid and their location is evolved in time inside
the grid by solving the equation d$\vec x$/d$t = \vec v$ with a second order accurate Euler 
integration scheme. A tracer particle is considered unbound if
$e_\mathrm{tot}>0$ and if its radial component of the velocity is
positive, steadily from a certain time until the end of the simulation.   
For each ejected tracer,
neutrino fluxes and mean energies as a function of time come from
the axisymmetric output provided by the hydrodynamic simulations,
see Sect.~\ref{subsec:hydro} and \cite{Perego.etal:2014}.

\subsection{Nucleosynthesis network}
\label{subsec:network}
For the post-processing of the tracers, a complete
nucleosynthesis network \citep{Winteler:2012,Winteler.etal:2012} is
employed. Over 5800 nuclei between the valley of stability and the
neutron drip line are considered, i.e. isotopes from H to Rg. The
reaction rates are taken from the compilation of
\cite{Rauscher.Thielemann:2000} for the Finite Range Droplet Model
(FRDM, \cite{Moeller.etal:1995}). Weak interaction rates including
neutrino absorption on nucleons are taken into account
\citep{Moeller.etal:2003,Froehlich:2006}. Furthermore, neutron capture for nuclei with $Z
\gtrsim 80$ and neutron-induced fission rates are given by
\cite{Panov.etal:2010} as well as $\beta$-delayed fission
probabilities from \cite{Panov.etal:2005}.

We perform nucleosynthesis calculations for over $17\,000$ ejected tracers
from the hydrodynamical simulation (see
Sect.~\ref{subsec:hydro-results}). Our computations start when the
temperature drops below $T = 10~\mathrm{GK}$. Then the initial
composition is determined by nuclear statistical equilibrium (NSE) and
is dominated by alpha particles, neutrons, and protons. NSE is assumed
to hold for $T \gtrsim 8~\mathrm{GK}$. Between 10~GK~$>T>$~8~GK, the
network evolves the weak reactions since they are not in equilibrium
and change the $Y_e$ accordingly. As soon as the temperature
undershoots the NSE threshold, the full network provides the
abundances. The longest trajectories were simulated until
$t_\mathrm{sim} \sim 200~\mathrm{ms}$, therefore we extrapolate them
following the prescription outlined in \cite{Korobkin.etal:2012}.

Moreover, the energy generation by the r-process is calculated and its
impact on the entropy is included \citep{Freiburghaus.etal:1999a}. 
As is common practice, we neglect the feedback of nuclear heating on the
density evolution. In the case of the dynamic ejecta, it has been shown that
this approximation is very good for nucleosynthesis calculations, although the
long-term hydrodynamical evolution of the ejecta is heavily
influenced\footnote{We expect that the impact on the nucleosynthesis in
neutrino-driven winds is small, but since the expansion time scales here are
different from the dynamic ejecta, this issue may need further scrutiny in the
future.}\citep{Rosswog.etal:2014}. The heating mainly
originates from $\beta$-decays and we assume that the energy is
roughly equally distributed between thermalizing electrons and photons, and
escaping neutrinos and photons \citep{Metzger.etal:2010b}.

\section{Nucleosynthesis}
\label{sec:results}

\subsection{Time and angle dependency}
\label{sec:angle-dependency}

The composition and amount of matter ejected in the neutrino-driven
wind depends on the temporal evolution of the disk and the fate of the 
compact central object. Here we consider a long-lived MNS. In the
following, we present abundances at various times after the merger and
for different latitudes to understand the angular distribution and
temporal evolution of ejected matter. The potential consequences for the mixing with
the dynamic ejecta and thus for the light curve are discussed in
Sects.~\ref{sec:mixing} and \ref{sec:em-signal}, respectively. 
In the following, $t_{\rm sim}$, refers to the time since merger.
When abundances are shown at $t_\mathrm{sim}$, we consider all tracers having become unbound until this
time. Therefore, at later times early ejecta are also included. 

\begin{figure}[!htb]
  \begin{center}
    \includegraphics[width=0.95\linewidth]{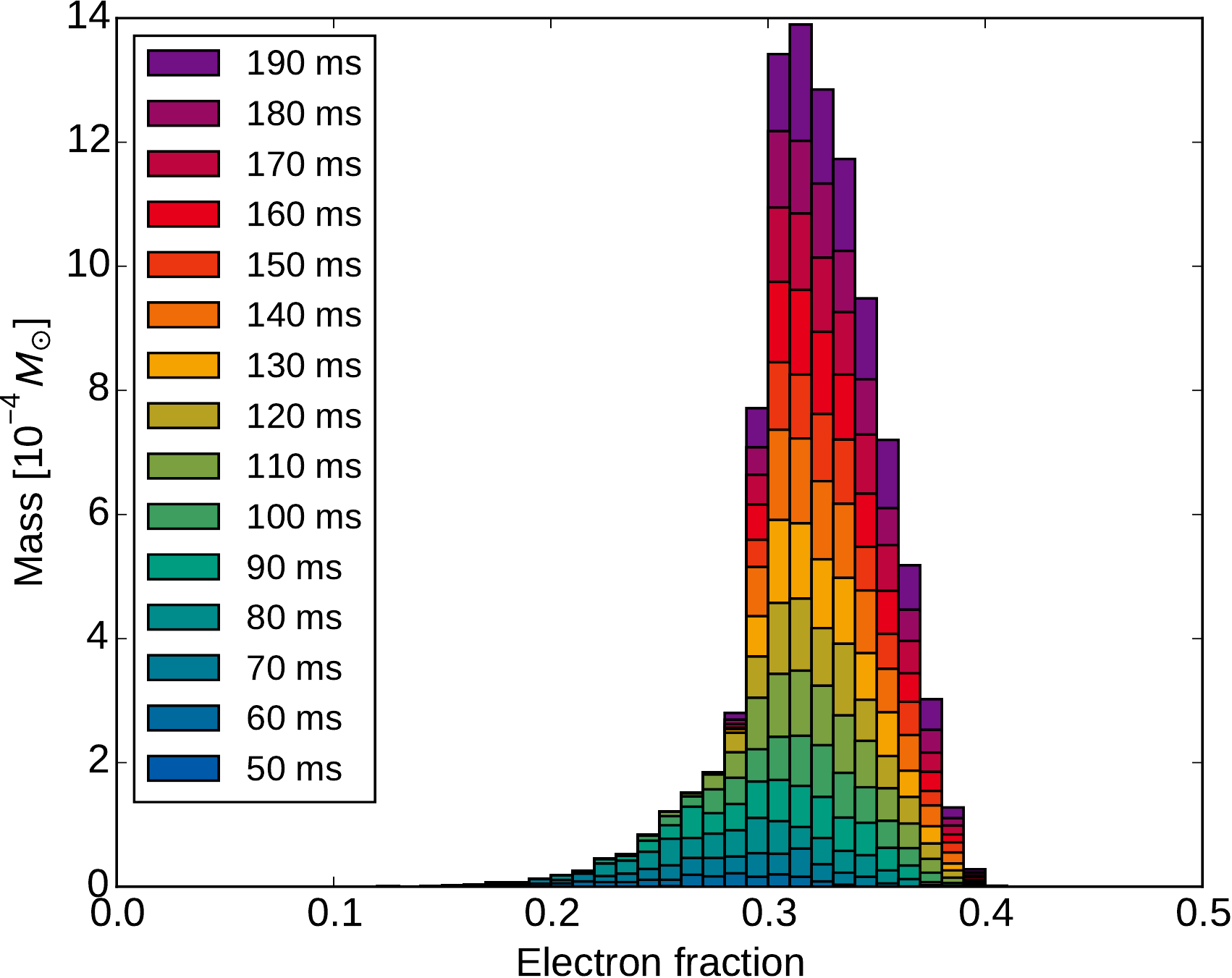}
  \end{center}
  \caption{Evolution of the electron fraction for ejected tracers. The
    $Y_{\rm e}$ distributions at various simulation times are represented by
    superposing bars with different colors.}
  \label{fig:tracers-ye-evolution}
\end{figure}

An idea of the amount of matter ejected and its potential composition can
be gained by exploring the dependence of the ejected mass on the electron
fraction and time. This evolution is shown in
Fig.~\ref{fig:tracers-ye-evolution}. Within the first $50$~ms only a 
marginal amount of mass with electron fraction $Y_{\rm e} \lesssim 0.3$ 
gets unbound. Until $t_\mathrm{sim} = 100$~ms 
approximately $2 \cdot 10^{-3} M_\odot$ are ejected with a central value 
of $Y_{\rm e} \approx 0.3$. After almost twice 
that time, i.e. $t_\mathrm{sim} = 190$~ms, the mass of unbound material 
reaches $9 \cdot 10^{-3} M_\odot$. These late-time ejecta has relatively 
high electron fractions (with a mean value between 0.3 and 0.35) since 
neutrinos have more time to interact with nucleons. Furthermore, the growth rate of
unbound mass by the late ejection of tracers remains approximately
constant from 110~ms on. The substantial mass accretion rate inside the accretion disk, 
amounting to a few tenths of solar masses per second, provides 
quasi steady-state conditions with neutrinos as primary cooling agent. 
As a consequence, the neutrino luminosities decrease only slowly over the expansion time scale
(cf. Fig.~10 in \cite{Perego.etal:2014}) and the interaction rates get close to 
equilibrium.

In the following, abundances are shown using either of these two
normalizations for a given nuclear mass $A$:
\begin{itemize}
\item Total mass: 
\begin{equation}
  \label{eq:total-massfractions}
  X_{\rm tot} (A) = \sum_{i=1}^N X_i (A) \cdot m_i,
\end{equation}
with $N$, $X_i$ and $m_i$ being the number of tracers considered, the mass fractions and the mass of one tracer
$i$, respectively. This quantity has mass units and thus allows to
compare different ejection channels to each other, as we will discuss in
Sect.~\ref{sec:mixing}.
\item Integrated abundances: 
\begin{equation}
  \label{eq:integrated-abundances}
  Y_{\rm int} (A) = \frac{\sum_{i=1}^N Y_i(A) \cdot m_i}{\sum_{i=1}^N m_i}. 
\end{equation}
Normalizing the abundances via 
Eq.~(\ref{eq:integrated-abundances}) allows to explore the average nucleosynthesis 
yields of a certain subclass of tracers, for instance tracers ejected into a 
particular solid angle. Additionally, the impact of single trajectories on the 
overall nucleosynthesis is directly evident from the comparison with the average 
abundance curve.
\end{itemize}

\begin{figure}[!htb]
  \begin{center}
    \includegraphics[width=0.95\linewidth]{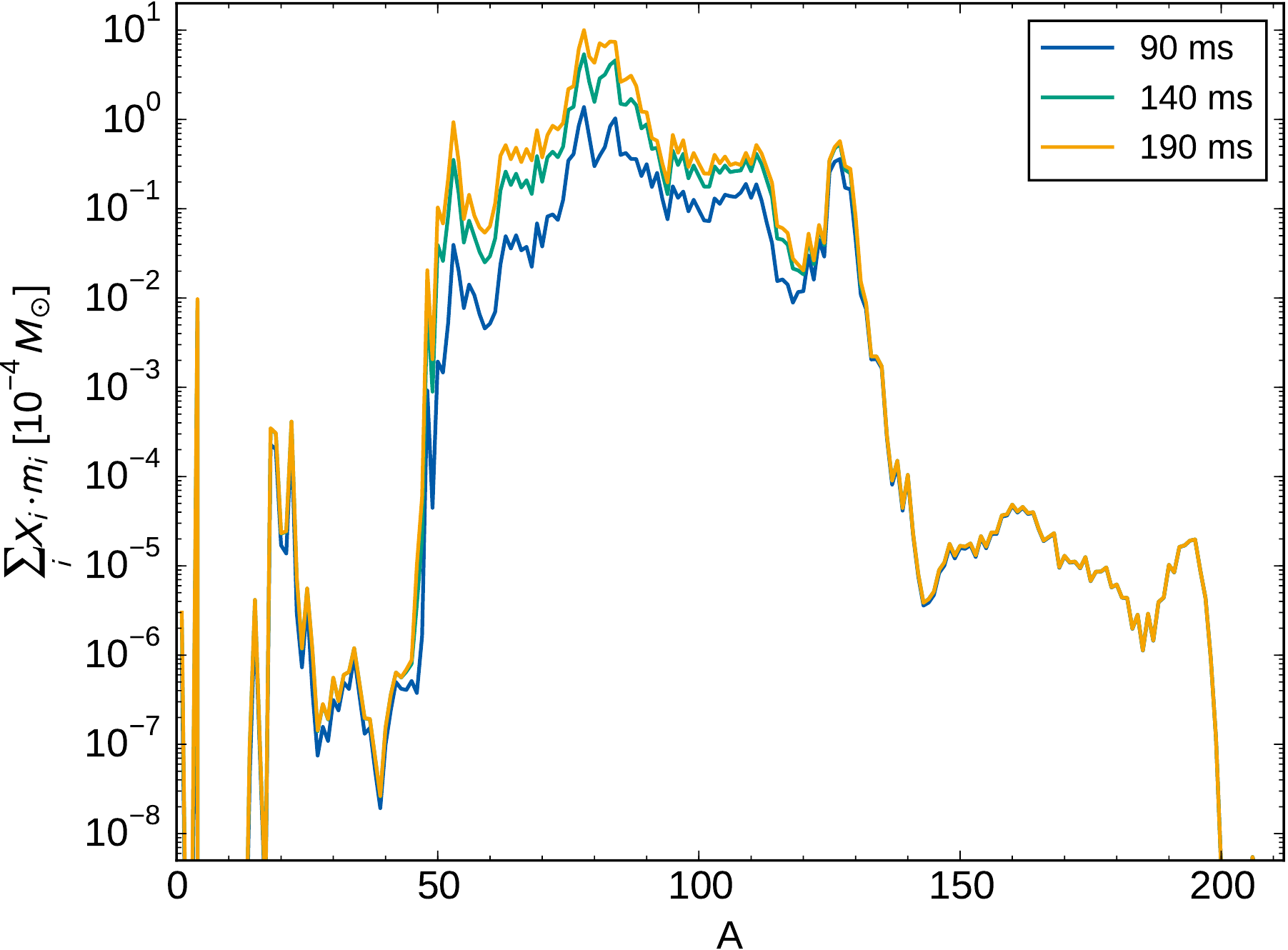}
  \end{center}
  \caption{Total final mass fractions times the mass of the
    ejecta until $t_\mathrm{sim} = 90~\mathrm{ms}$, $140~\mathrm{ms}$ and
    $190~\mathrm{ms}$. Early neutron-rich ejecta produce heavy
    r-nuclei with $A \gtrsim 130$, leading to overlapping abundances
    in this region. Later ejecta contribute strongly to the lighter
    heavy elements (note the logarithmic ordinate).}
  \label{fig:total-nucleosynthesis-Ym}
\end{figure}

Figure~\ref{fig:total-nucleosynthesis-Ym} shows total ejected
masses for simulation times $t_\mathrm{sim} = 90~\mathrm{ms}$,
$140~\mathrm{ms}$ and $190~\mathrm{ms}$. The heavy nuclei beyond the
second r-process peak ($A \sim 130$) are produced by early, very
neutron-rich ejecta, as indicated by the overlap of the three curves
at different simulation times. Later on, no heavy r-process elements
$(A \gtrsim 130)$ are produced any more. There is a threshold
$Y_{\rm e} \sim 0.25$ below which heavy r-process elements (i.e., beyond the
second peak at $A=130$) can be synthesized (see also 
\cite{Kasen.etal:2015}). As time passes, a substantial amount of tracers
contributing to nuclei with $A \lesssim 120$ becomes unbound. This
evolution is in fact a direct consequence of the trend of $Y_{\rm e}$
presented in Fig.~\ref{fig:tracers-ye-evolution}. 

\begin{figure}[!htb]
  \begin{center}
    \includegraphics[width=0.95\linewidth]{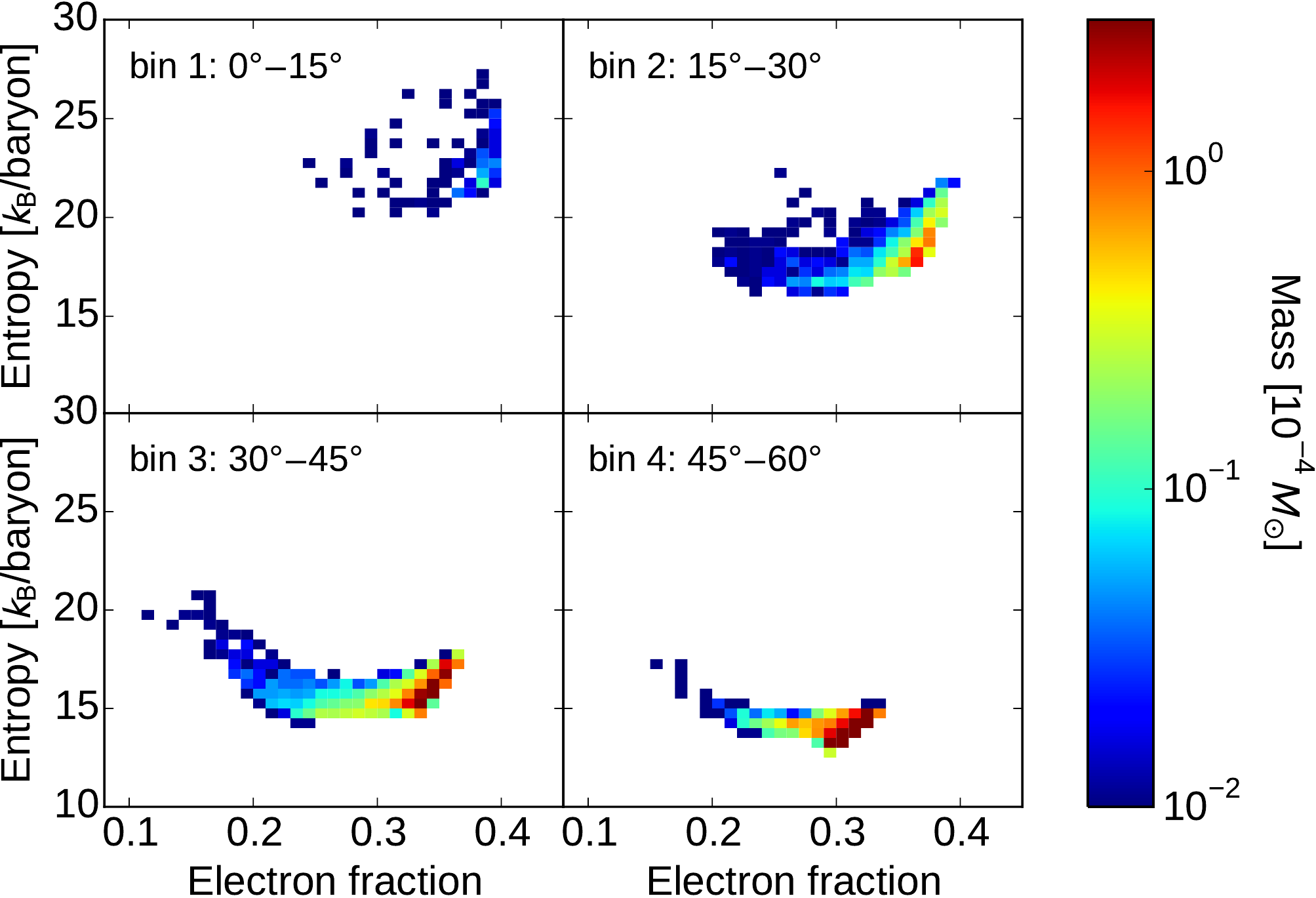}
  \end{center}
  \caption{Distribution of the unbound tracers in the $Y_{\rm e} - s$ plane
    at a sphere with a radius of 750~km. We present the logarithmic
    scale of the total mass for all tracers in a certain angular bin of the
    $Y_{\rm e} - s$ plane with a color code. The panels correspond to the
    angle intervals defined in Fig.~\ref{fig:simulation-rho-v}.}
  \label{fig:tracers-ye-S}
\end{figure}

\begin{figure}[!htb]
  \begin{center}
    \includegraphics[width=0.95\linewidth]{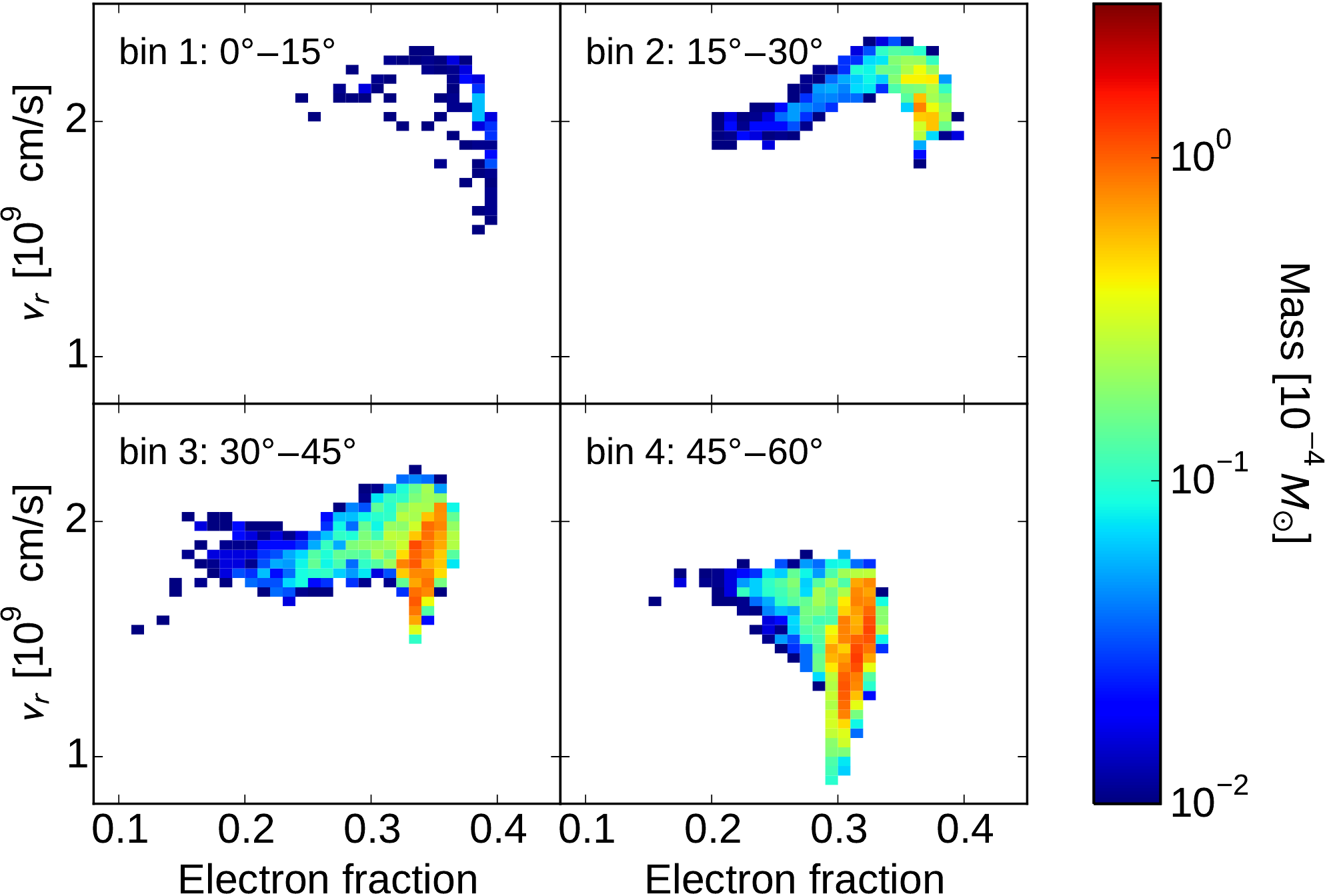}
  \end{center}
  \caption{Same as Fig.~\ref{fig:tracers-ye-S}, but for the $Y_{\rm e} -
    v_r$ plane.}
  \label{fig:tracers-ye-v}
\end{figure}

In addition, the nucleosynthesis has an angular dependency. We
divide the neutrino-driven wind into four regions above and below the
disk by setting cuts on the polar angle and investigate the angular
dependence of the ejecta. Each of the cuts has a width of 
$\Delta \theta = 15^\circ$, hence the whole neutrino-driven wind is 
captured within $0^\circ \leq \theta \leq 60^\circ$ (as a convention we define 
$\theta = 0^\circ$ at the poles). These four regions or bins
are indicated in Fig.~\ref{fig:tracer-xz} with white dashed lines and are labeled with the number of
the bin. The properties of the tracers ejected until $t_\mathrm{sim}=190$~ms in each
of the angular bins are shown in Figs.~\ref{fig:tracers-ye-S} and
\ref{fig:tracers-ye-v}, when passing a sphere with a radius of 750~km. We select the nucleosynthesis-relevant 
quantities: electron fraction $Y_\mathrm{e}$, entropy per baryon $s$ and radial 
velocity $v_r$. The major part of the cumulative mass is approximately 
equally distributed to the two angular bins close to the accretion disk. On the
contrary, the two bins in the polar region contain 15\%~$-$~20\% of the
total cumulative mass. While the entropy of the unbound tracers ranges from
10~$k_\mathrm{B}$/baryon to 30~$k_\mathrm{B}$/baryon
(Fig.~\ref{fig:tracers-ye-S}), it is still very low and has therefore 
little impact on the nucleosynthesis. The radial velocity provides a measure for the dynamical 
time-scale and is constrained to $0.04c - 0.08c$ 
(Fig.~\ref{fig:tracers-ye-v}). In the neutrino-driven wind
the nucleosynthesis is most sensitive to the electron fraction 
distribution which varies for every angular region. As a general trend,
the average electron fraction decreases as a function of the angle and
reaches values down to 0.3 for the two zones closest to the disk. Both 
of these bins also contain extreme cases with very neutron-rich conditions 
of $Y_\mathrm{e} < 0.2$.


\begin{figure}[!htb]
  \begin{center}
    \includegraphics[width=0.95\linewidth]{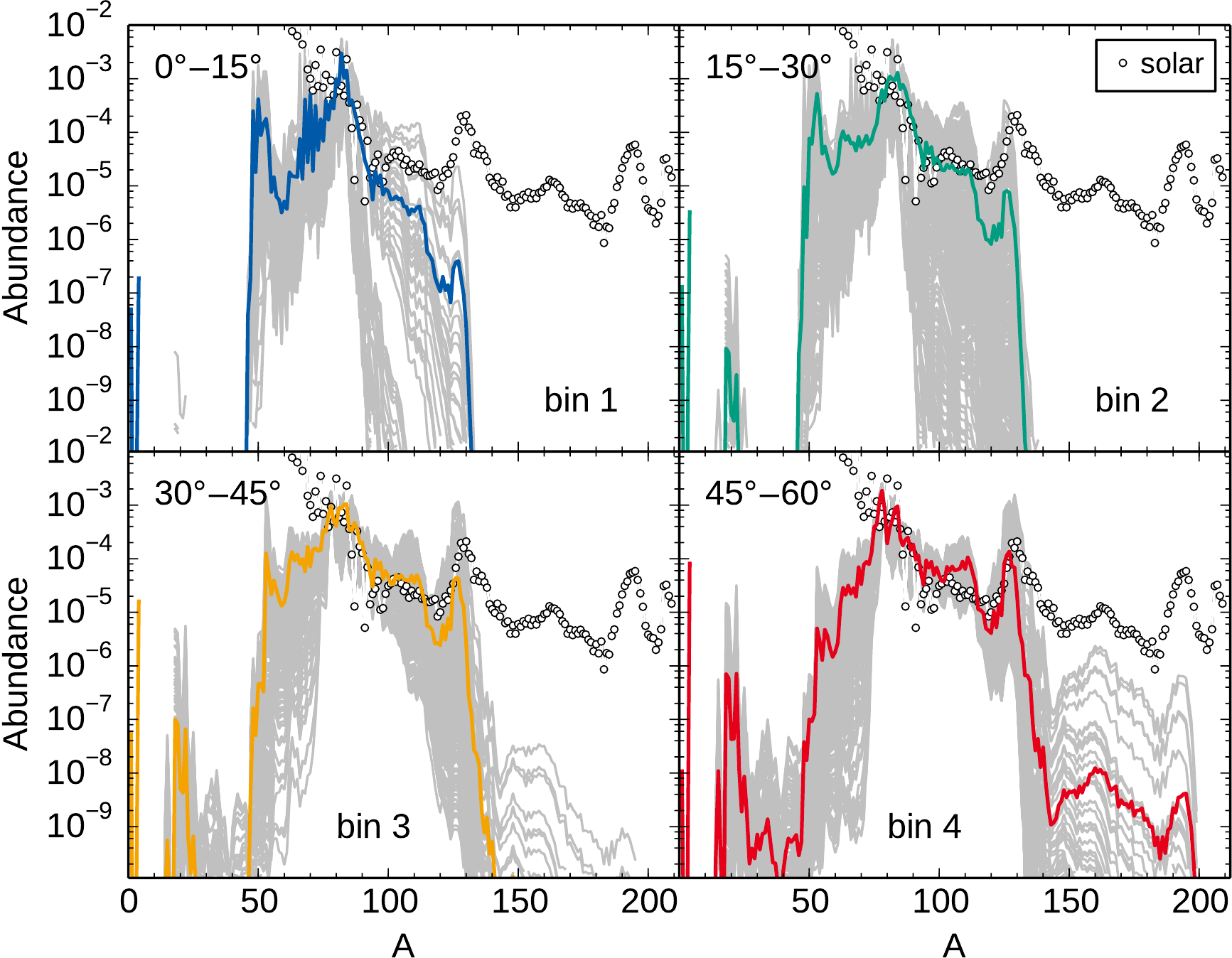}
  \end{center}
  \caption{Individual abundances and integrated final abundances for ejecta until
    $t_\mathrm{sim} = 140~\mathrm{ms}$ in each of the four considered angular bins. Thick colored lines mark the
    average yields for each angular region. Note that we apply the same
    color code consistently in all following figures. Moreover, thin
    gray lines denote abundances from individual trajectories to hint
    the variety of the nucleosynthesis. Solar abundances are shown
    with dots for comparison.}
  \label{fig:integrated-nucleosynthesis-140ms} 
\end{figure}

Figure~\ref{fig:integrated-nucleosynthesis-140ms} shows the resulting
integrated abundances for tracers ejected at $t_\mathrm{sim} = 140~\mathrm{ms}$. 
All yields are presented in comparison to the solar abundances 
(dots, \cite{Lodders:2003}) for one angular bin in every panel. 
The thin gray lines represent individual tracers, while averages are 
indicated by a solid thick line in every panel. The patterns in
Fig.~\ref{fig:integrated-nucleosynthesis-140ms} reveal significant
differences in abundances for distinct latitudes. We find that the
first r-process peak $(A \sim 80)$ forms for each angular region,
whereas the second abundance peak ($A \sim 130$) is only reached in
bin~3 and bin~4. In particular, the angular zone closest to the disk,
i.e., bin~4, successfully attains elemental abundances close to the
solar system values up to the second abundance peak. When moving to
lower polar angles (bin~1), less heavy elements are synthesized.

\begin{figure}[!htb]
  \begin{center}
    \includegraphics[width=0.95\linewidth]{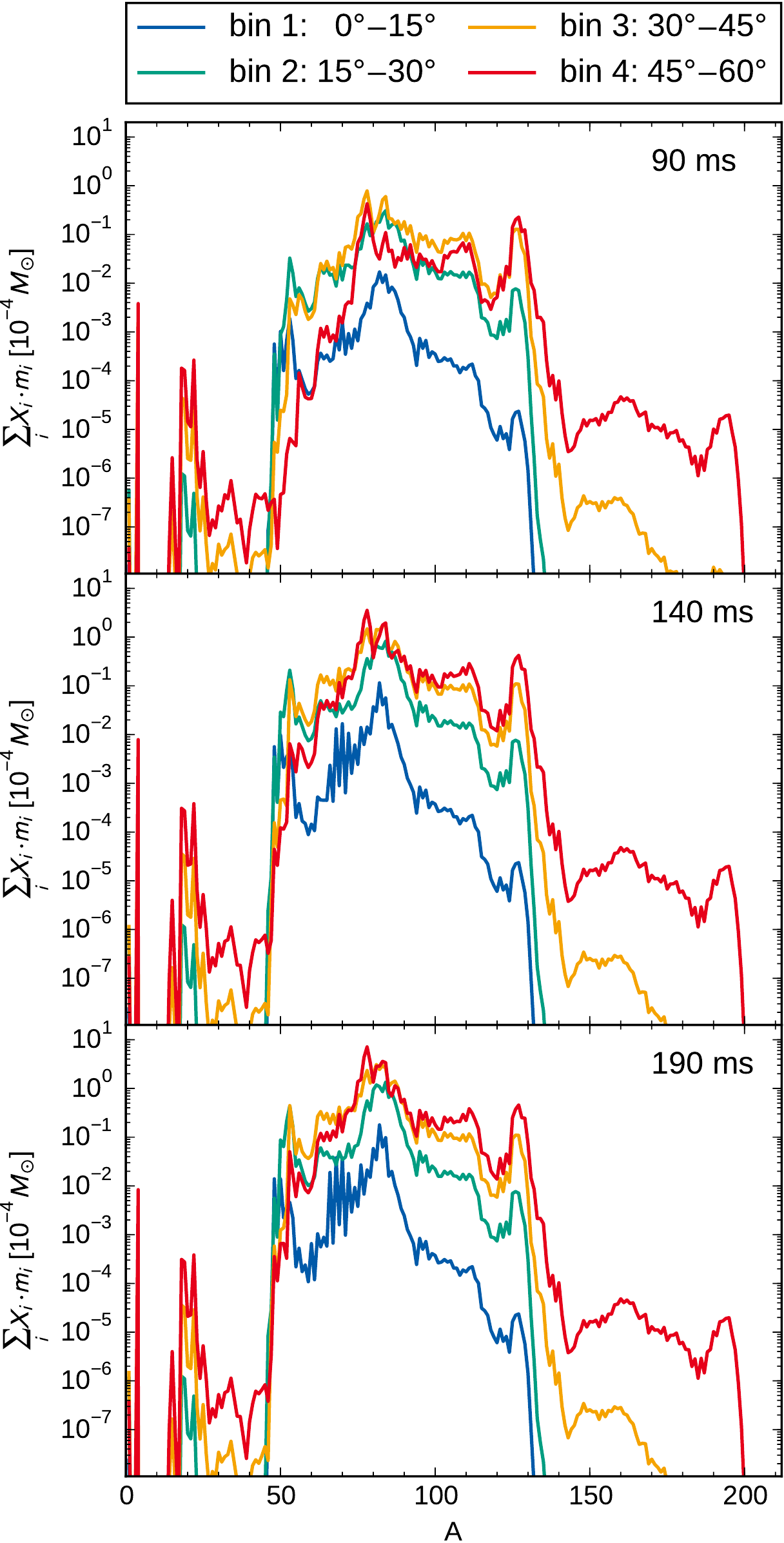}
  \end{center}
  \caption{Total final masses. The panels show the yields
    for tracers ejected until $t_\mathrm{sim} = 90~\mathrm{ms}$,
    $140~\mathrm{ms}$ and $190~\mathrm{ms}$ from top to bottom. We use
    the same color scheme as in
    Fig.~\ref{fig:integrated-nucleosynthesis-140ms} to identify the
    different angular bins.}
  \label{fig:Ym-90-190ms}
\end{figure}

To gain further insights into the dependence on time, it is
instructive to compare the trends of the angular bins at various
times. Total masses as a function of mass number $A$
(Eq.~(\ref{eq:total-massfractions})) for unbound tracers at
$t_\mathrm{sim} = 90~\mathrm{ms}$, $140~\mathrm{ms}$ and
$190~\mathrm{ms}$ are presented in Fig.~\ref{fig:Ym-90-190ms}. At
early times $t_\mathrm{sim} \lesssim 90~\mathrm{ms}$, elements up to
$A \sim 120$ are produced predominantly by the ejecta in bin~3, while
only bin~4 synthesizes nuclei within the vicinity of the third
abundance peak ($A \sim 195$). The yields resulting from 
both bin~3 and bin~4 clearly dominate the abundances, as is
expected from the distribution of the mass and the electron fraction
for the tracers among the four bins (cf. Figs.~\ref{fig:tracers-ye-S} 
and \ref{fig:tracers-ye-v}). In contrast, bin~1 and bin~2 give 
marginal contributions. With time, increasing yields for 
lighter heavy nuclei ($A \lesssim 120$) 
(cf. Fig.~\ref{fig:total-nucleosynthesis-Ym}) are present in all
considered angular regions.

\begin{figure}[!htb]
  \begin{center}
    \includegraphics[width=0.95\linewidth]{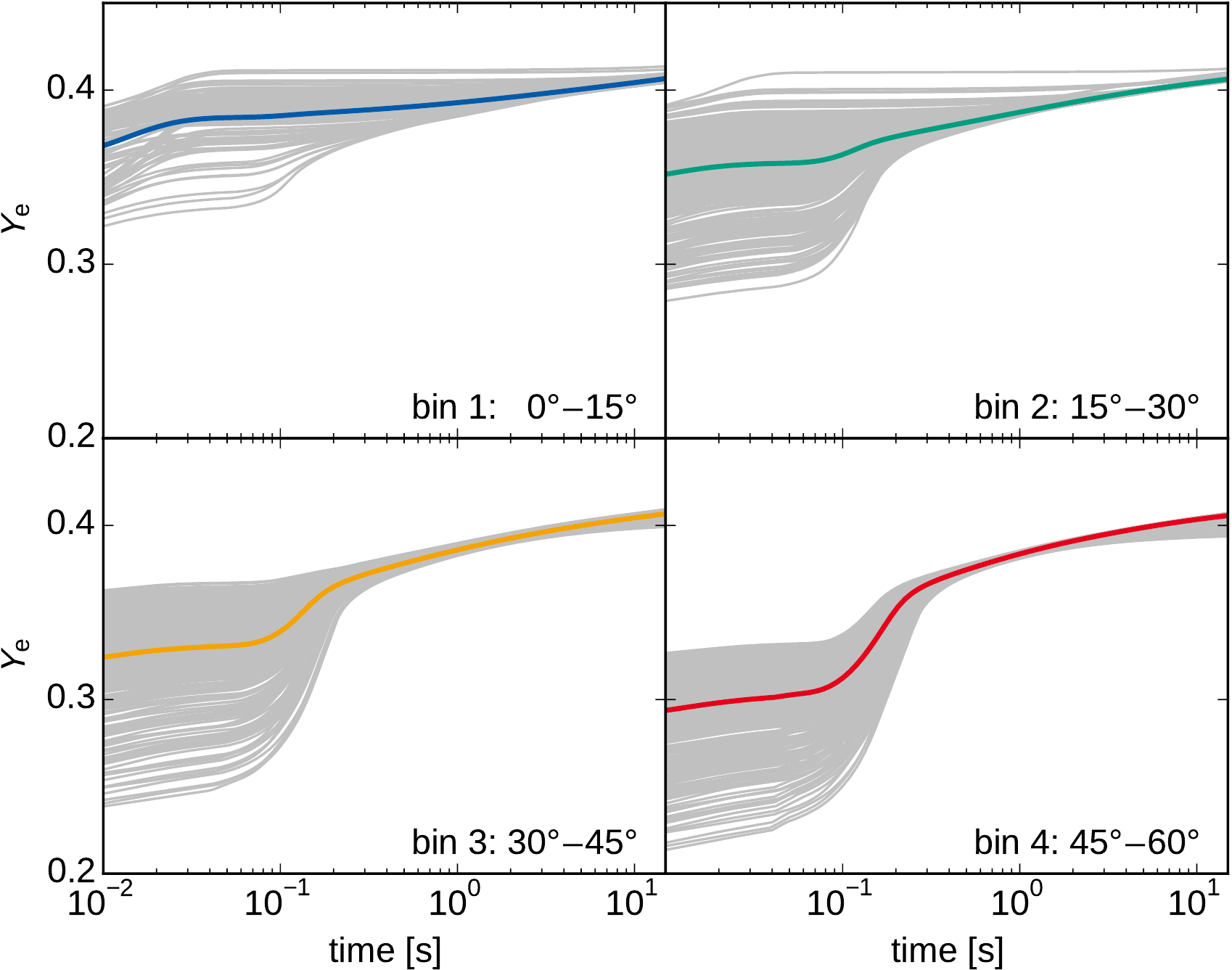}
  \end{center}
  \caption{Temporal evolution of the electron fraction for tracers
    that become unbound until $t_\mathrm{sim} = 140~\mathrm{ms}$. The
    solid thick lines are averages $\langle Y_{\rm e} \rangle$ for every
    angular bin as introduced in
    Fig.~\ref{fig:integrated-nucleosynthesis-140ms}. Thin gray lines
    represent individual tracers.}
  \label{fig:ye-evolution}
\end{figure}

Let us take a closer look at the relevant physical quantities
characterizing the nucleosynthesis. Figure~\ref{fig:ye-evolution}
shows the evolution of individual electron fractions for tracers 
unbound at 140~ms (gray, thin lines), along with the
corresponding average curves (colored, thick
lines). Note that the elapsed time of a tracer is defined with 
respect to the moment when it gets out of NSE in the network calculations. 
The individual tracers provide information about the spread in the
initial $Y_{\rm e}$ (i.e. at $T = 10~\mathrm{GK}$), also visible in
Figs.~\ref{fig:tracers-ye-S}~and~\ref{fig:tracers-ye-v}. The slight increase
of the electron fraction shortly after the beginning of the
calculations is due to neutrino absorption on nucleons, whereas the
steep rise at around $t \simeq 10^{-1}~\mathrm{s}$ is caused by
$\beta$-decays.

The electron fraction in the neutrino-driven wind depends on the
competition between the expansion and the weak equilibrium time
scales, on the ratio between the $\nu_e$ and $\bar{\nu}_e$
luminosities, as well as on the neutrino mean energies (e.g.,
\cite{Qian.Woosley:1996}). As we mentioned in
Sect.~\ref{subsec:hydro}, the larger luminosities predicted by GR simulations
can lead to a faster
evolution towards the equilibrium $Y_{\rm e}$. Therefore, the very
small amount of heavy elements ($A>130$) that are produced in the
initial phase of our Newtonian simulation may not be present in a GR
simulation. In addition to the GR effects, an accurate calculation
of $Y_{\rm e}$ would require the usage of a detailed neutrino transport
scheme, which is presently computationally prohibitive in long-time
three dimensional simulations, without global symmetries. Assuming
uncertainties of the order of 20\% for the neutrino luminosities and
of 10\% for the mean energies\footnote{This estimate is consistent
with the quantitative differences spotted between our ASL scheme
results and the ones obtained using a multigroup flux limited
diffusion transport in axisymmetric simulations \citep{Dessart.etal:2009}.}, 
we estimate a potential uncertainty of $\sim 15 \%$ on the values of the
equilibrium electron fraction. Given the broad range of $Y_{\rm e}$ obtained
in the ejecta (Figs.~\ref{fig:tracers-ye-S}, \ref{fig:tracers-ye-v} and
\ref{fig:ye-evolution}), we consider this uncertainty as being important to be
mentioned, but not crucial for our analysis.

\begin{figure}[!htb]
  \begin{center}
    \includegraphics[width=0.95\linewidth]{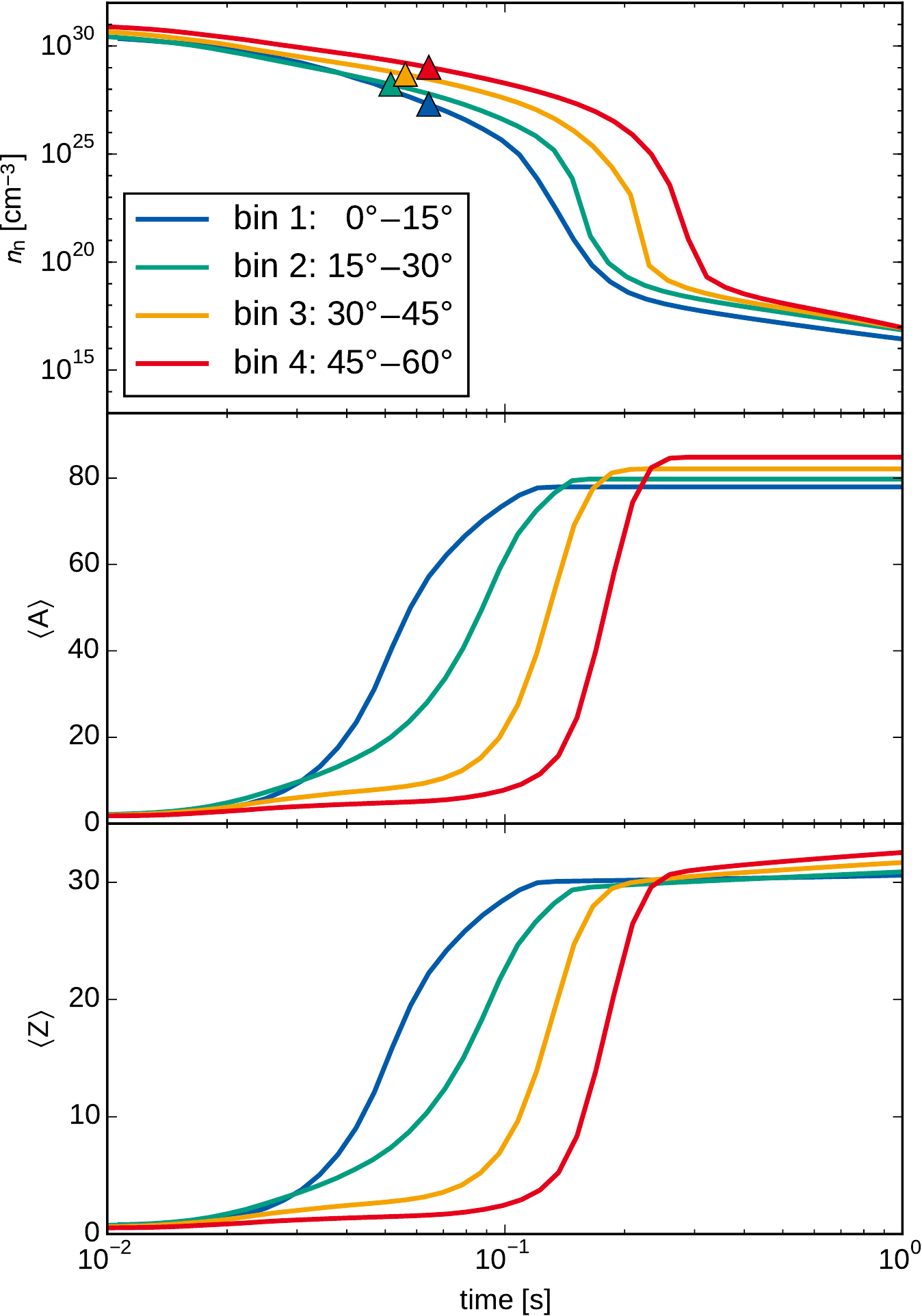}
  \end{center}
  \caption{Evolution of the neutron density $n_{\rm n}$ (top), average mass
    number $\langle A \rangle$ (middle) and average proton number
    $\langle Z \rangle$ (bottom). The solid lines follow the mean
    values of the four angular bins (see also
    Fig.~\ref{fig:ye-evolution}). Triangles indicate times, when the
    temperature reaches $T = 3~\mathrm{GK}$, i.e. the onset of the
    r-process.}
  \label{fig:Nn-Abar-Zbar}
\end{figure}

In order to examine the production of heavy elements, we present the
evolution of neutron density $n_n$, the average mass number $\langle A
\rangle$ and the average proton number $\langle Z \rangle$ in
Fig.~\ref{fig:Nn-Abar-Zbar} for different angular regions. At the
beginning, the neutron density is larger than
$10^{30}~\mathrm{cm}^{-3}$ and the composition is dominated by alpha
particles and neutrons. The onset of the r-process nucleosynthesis is
triggered when the temperature decreases to $\sim 3$~GK
\cite[marked by triangles in the top panel of Fig.~\ref{fig:Nn-Abar-Zbar}; 
see, e.g., the dependence of r-process efficiency as a function of entropy, 
electron fraction and expansion time scale,][]{Hoffman.etal:1997,Freiburghaus.etal:1999b}. For $t \simeq (1.5 - 3.0) \cdot
10^{-1}~\mathrm{s}$, most of the neutrons are consumed as indicated by
the rapid neutron density drop. After this, matter beta-decays to
stability and this leads to an increase of $Y_{\rm e}$ (see
Fig.~\ref{fig:ye-evolution}). Then, the mass number stagnates,
since no more heavy elements are formed (see middle panel of
Fig.~\ref{fig:Nn-Abar-Zbar}). No fission takes place, resulting in
the monotonic evolution of mass and proton number. The higher neutron
density in bin~4 favors the build-up of heavier nuclei compared to the
other regions. With decreasing polar angle the neutrons are consumed
earlier, as their initial density is smaller and the expansion velocity 
of the corresponding tracers is larger.

\subsection{Neutrino-driven wind and dynamic ejecta}
\label{sec:mixing}
The neutrino-driven wind is only one of three nucleosynthesis-relevant
ejecta in neutron star mergers. In order to have the complete
nucleosynthesis picture, one would need to follow the evolution of the
dynamic and disk (neutrino- and viscous-driven) ejecta, and investigate
how they mix. However, this is currently too complicated to be 
studied in a single simulation due to the different
time-scales and physics involved. The dynamic ejecta expand very fast from the
beginning. While neutrinos may be initially important for the $Y_{\rm e}$
their impact is insignificant when the disk becomes transparent
\citep{Fernandez.Metzger:2013,Metzger.Fernandez:2014,Just.etal:2015a}.
Nevertheless, mixing of the different types of ejecta can occur once
the unbound matter becomes considerably decelerated by the ambient
medium.  Where this happens depends on the merger location with
respect to the galaxy. Simple estimates indicate that the time scale
for being noticeably decelerated is by far greater than years, see,
for example, the discussion in Sect.~2.1 of \cite{Rosswog.etal:2014}).

In the first study of the mixing between the disk and dynamic ejecta, \cite{Fernandez.etal:2015}
showed that there is no significant mixing between these ejecta. This
study focused on neutron star black hole mergers, where neutrino-driven
ejecta are less important as it was discussed also by
\cite{Fernandez.Metzger:2013} compared to \cite{Metzger.Fernandez:2014}. However, their
conclusion may also apply to our results that are based on a
long-lived MNS. From the
nucleosynthesis perspective, \cite{Just.etal:2015a} have also investigated the
mixing of disk and dynamic ejecta. They have modeled the long-time
evolution of a black hole surrounded by a disk where matter is
neutrino- and viscous-driven ejected. In a post-processing step, the
nucleosynthesis of such simulations was added to the dynamic ejecta
from a merger simulation \citep{Bauswein.etal:2013}. 

We combine the nucleosynthesis of the neutrino-driven wind presented 
in Sect.~\ref{sec:angle-dependency} with	the dynamic ejecta computed by
\cite{Korobkin.etal:2012}. Although the ejecta masses show a
substantial spread depending on the parameters of the merging binary
system \citep{Rosswog:2013}, the resulting abundance patterns have
been found to be practically identical
\citep{Korobkin.etal:2012}. 
In contrast to previous studies (i.e., \cite{Fernandez.Metzger:2013, Just.etal:2015a})
with a black hole as central object, we assume a long-lived
MNS. This has strong consequences on the amount and properties
(i.e., $Y_{\rm e}$) of the matter ejected by neutrinos as discussed by
\cite{Perego.etal:2014, Metzger.Fernandez:2014, Kasen.etal:2015}. 

\begin{figure}[!htb]
  \begin{center}
    \includegraphics[width=0.95\linewidth]{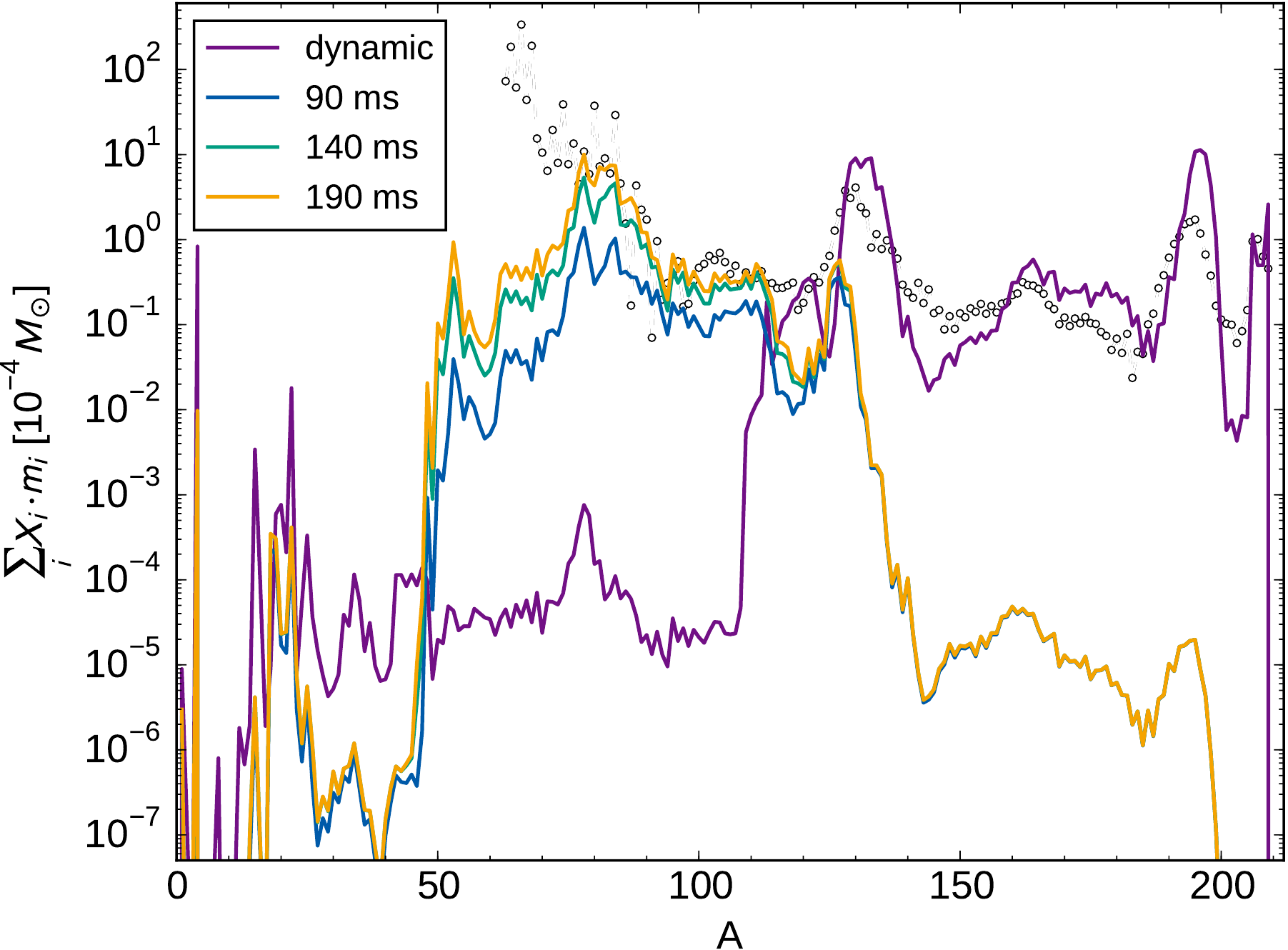}
  \end{center}
  \caption{Comparison of the nucleosynthesis yields for neutrino-driven wind (cf. Fig.~\ref{fig:total-nucleosynthesis-Ym})
    and dynamic ejecta (labeled with dynamic). While the dynamic ejecta produce very heavy nuclei, 
    the neutrino-driven wind complements its abundances by producing the
    lighter heavy elements from the first to the second r-process peak.}
  \label{fig:disk-vs-dynamic-ejecta-A}
\end{figure}

The comparison between dynamic and wind ejecta is given in
Fig.~\ref{fig:disk-vs-dynamic-ejecta-A} showing the total ejected
masses (Eq.~(\ref{eq:total-massfractions})) as a function of $A$ for
both contributions. Note that the approach here differs from the one
in the previous sections, in which we treated the different times
$t_{\rm sim}$ to be merely snapshots of the neutrino-driven wind. Now,
we assume that the MNS collapses after $t_{\rm sim}$, terminating the
ejection of further material in the wind. Three different simulations
times are considered, $t_\mathrm{sim}=$~90~ms, 140~ms, and
190~ms. This comparison indicates that the wind ejecta complement the
dynamic ejecta by producing elements below the second peak. 
We note that if the dynamic ejecta could produce the full 
  r-process, as suggested by	recent GR simulations 
  \citep{Wanajo.etal:2014,Sekiguchi.etal:2015} and by parametric 
  studies \citep{Goriely.etal:2015}, the combination of the dynamic 
  and wind nucleosynthesis would again allow the production of 
  lighter heavy elements ($A < 130$)	in addition to the heavy 
  r-process. In the latter case, the wind ejecta would increase only
  the relative contribution of the lighter heavy elements.
Moreover, the amount of wind ejecta becomes comparable to the dynamic
ejecta if the MNS survives long enough ($t_\mathrm{sim} \gtrsim
$~140~ms in our model). These ejecta amounts will be further enhanced
by viscous ejecta that can produce r-process elements from the first
to the third peak \citep{Just.etal:2015a}. If the three ejecta
completely mix, the wind contribution may still lead to variations in
the abundances below $A=130$ (i.e., $Z<50$). Such variations are also
expected from observations of the oldest stars (see
\cite{Sneden.etal:2008} for a review). The size of the observed
variations could help to constrain different contributions of the three
ejecta types to the r-process abundances.

\begin{figure}[!htb]
  \begin{center}
    \includegraphics[width=0.95\linewidth]{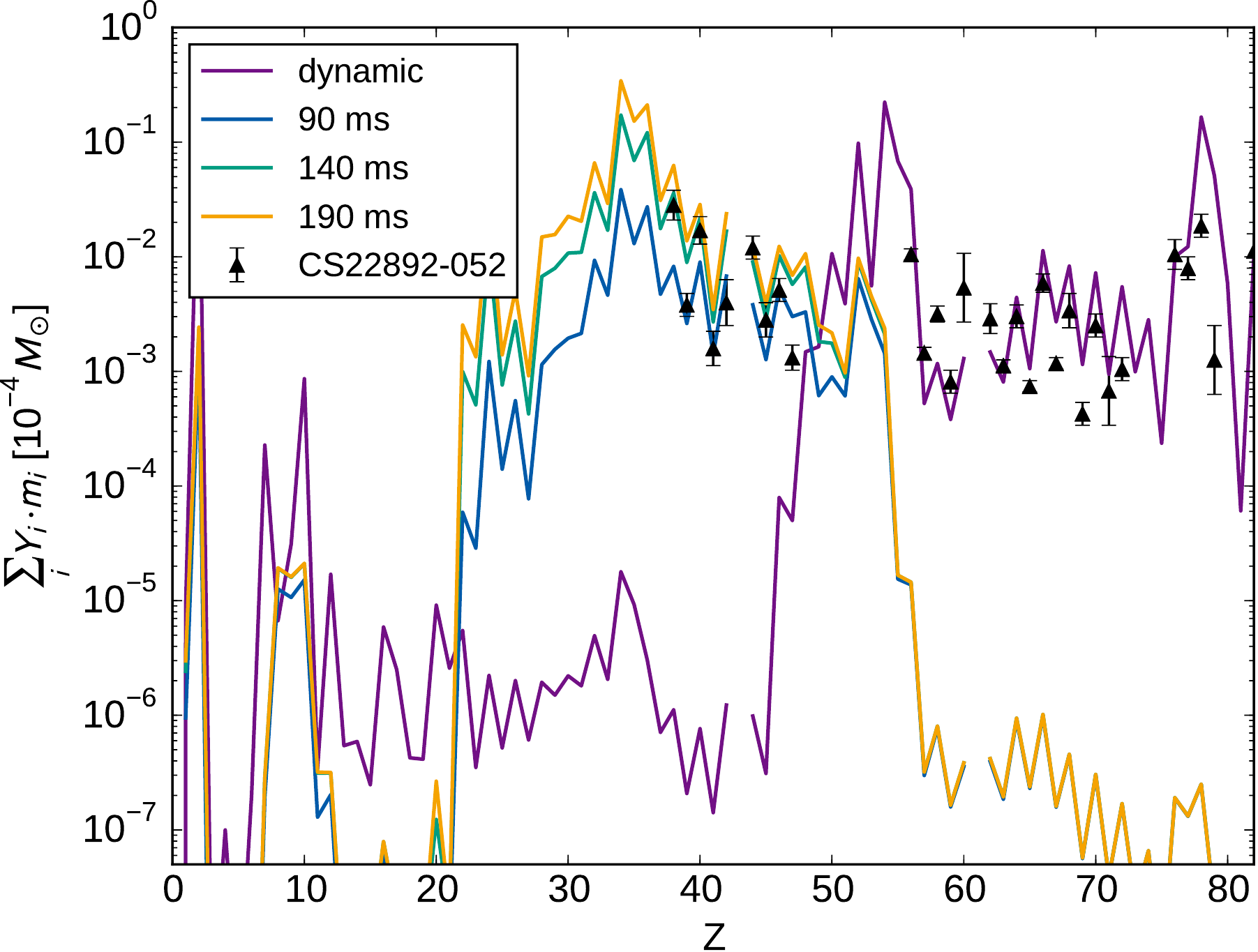}
    \includegraphics[width=0.95\linewidth]{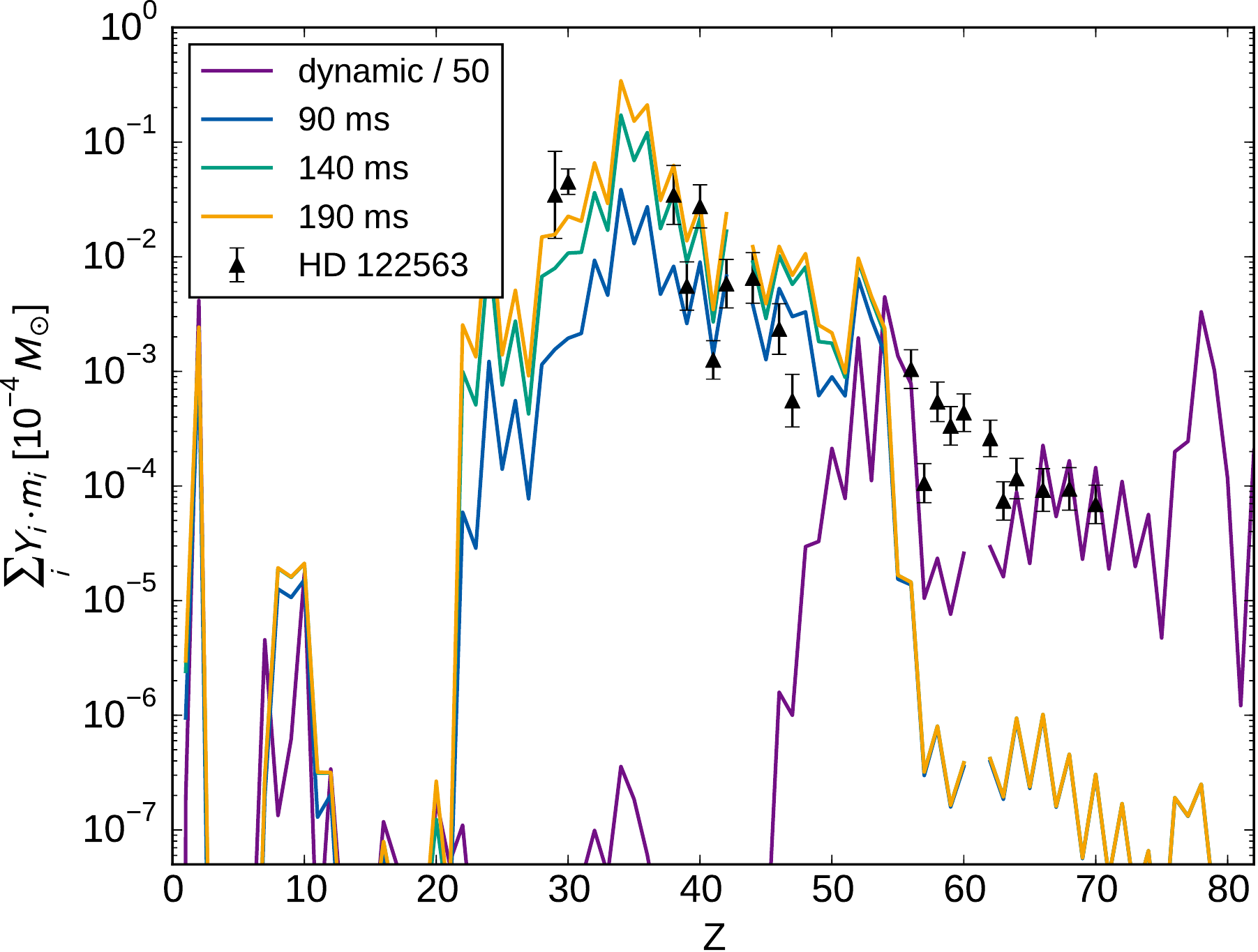}
  \end{center}
  \caption{Top: combined elemental abundances of neutrino-driven wind
    and dynamic ejecta in comparison to the yields of the metal-poor
    star CS22892-052 \citep{Sneden.etal:2003}. Abundances are weighted
    with the ejecta mass to consistently add the two
    contributions. Bottom: comparison of the combined yields with the
    ones of the HD stars
    \citep{Wallerstein.etal:1963,Honda.etal:2004,Honda.etal:2006}.
    Note that the dynamic ejecta are scaled down by a factor of 50.}
  \label{fig:mix-Z}
\end{figure}

The oldest observed stars were formed in an interstellar medium
polluted by only one or few nucleosynthesis events. Therefore, their
atmospheres present a unique fingerprint for the
r-process. Observations indicate that there are at least two types of
abundance patterns among the oldest r-process stars (see, e.g.,
\cite{Sneden.etal:2008,Qian.Wasserburg:2007}): 1) stars with high enhancement of
heavy elements ($Z>50$) present a robust pattern for those and some
variations for the lighter heavy elements ($Z<50$); 2) stars with low
enhancement of heavy r-process. Two stars are typically identified
with these trends: 1) CS~22829-052 \citep{Sneden.etal:2003} and 2) HD~122563
\citep{Wallerstein.etal:1963,Honda.etal:2004,Honda.etal:2006}. Recently,
\cite{Hansen.etal:2014} have shown that this second kind of pattern can be
explained by the superposition of two components: an H-component
producing the heavy r-process elements (and maybe also lighter ones)
and an L-component contributing only below $Z=50$. In neutron star
mergers, the dynamic and viscous ejecta can account for the
H-component while the wind ejecta would be the L-component. Matter
with a perfect mixing of the three ejecta will lead to a pattern like
in CS~22829-052. This is shown in the upper panel of
Fig.~\ref{fig:mix-Z}, where the combination of wind (L-component) and
dynamic (H-component) ejecta reproduce the observed abundances. 
The differences around the third peak are due to nuclear physics
input as shown in e.g.~\cite{Eichler.etal:2015}. If the mixing is not perfect
and the wind ejecta combines only with a small amount of dynamic
and/or viscous ejecta, then one can explain the trend observed in
HD~122563 (see also \cite{Just.etal:2015a}). This is shown in the
bottom panel of Fig.~\ref{fig:mix-Z} where the dynamic ejecta has been
reduced by a factor of 50. In this case, one would expect to have more
variability in the observations depending on the amount of the
H-component that is mixed. Indeed observations show variability for
$Z>50$ in these kind of stars \citep{Roederer.etal:2010,
Hansen.etal:2014}.

On the time scale of star formation, it is likely that all the ejecta
have efficiently mixed, even if the wind expands perpendicular to the
disk and the other two are more homogeneously distributed or stay in
the disk plane. Do we expect stars to be polluted by a neutron star
merger preferentially only receiving axial ejecta rather than disk ejecta? Or do
Honda-like abundances (L-component) come from different sources 
(e.g. core-collapse supernovae with slightly neutron-rich conditions)?
More investigation of the long-term mixing is
necessary to understand the contribution of neutron star mergers to
the different patterns found in the oldest observed stars. 
However, the dynamic and wind ejecta will not mix before they are substantially 
decelerated by the dilute ambient medium. For example, assuming a particle 
density of $\sim 1$~cm$^{-3}$ for the ambient medium,
the estimated deceleration time scale is of the order of several years.
The dynamic ejecta expand earlier and with high initial velocities 
of $\sim 0.1c$. The neutrino-driven wind, in contrast, gets unbound after 
neutrinos have deposited enough energy. Therefore, the wind matter is 
ejected later and eventually reaches asymptotic velocities of up to $\sim 0.08c$. The 
different ejection times and velocities make it possible to observe a kilonova 
signal from different parts of the ejecta with different composition, as 
we discuss in the next section.

\subsection{The electromagnetic signal: a semi-analytical model}
\label{sec:em-signal}

\begin{figure}[!htb]
\begin{center}
  \includegraphics[width=0.95\linewidth]{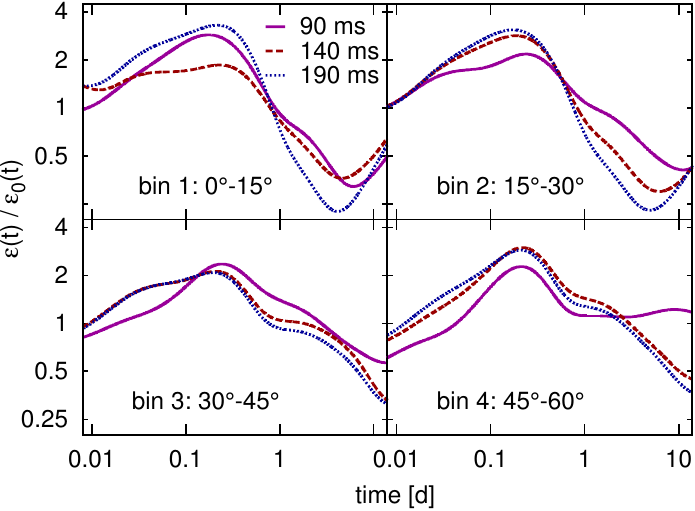} 
\end{center}
\caption{Radioactive heating rate, calculated with the nucleosynthesis network
for a representative tracer of the neutrino-driven wind in each angular bin, 
normalized to the fitted expression $\dot{\epsilon}_0(t)=10^{10}t^{-1.3}_{\rm d}\;{\rm
erg}\;{\rm g}^{-1}{\rm s}^{-1}$ (see, e.g., \cite{Metzger.etal:2010b,Korobkin.etal:2012}). In all cases the heating rate exhibits an excess
by a factor of $\sim2.5$ at $t\sim4$~h.
}
\label{fig:heating-rates}
\end{figure}

The nucleosynthesis network allows us to compute radioactive heating
rates for the wind outflow. Fig.~\ref{fig:heating-rates} shows the heating
rates for different bins, normalized to
$\dot{\epsilon}_0(t)=10^{10}t^{-1.3}_{\rm d}\;
 {\rm erg}\;{\rm g}^{-1}{\rm s}^{-1}$ with time $t_{\rm d}$ in units of days \citep{Metzger.etal:2010b,Korobkin.etal:2012}.
It is interesting to point out that all the normalized heating rates show
considerable excess at $t\sim4$~h.
This excess is caused by decay of radioactive nuclei in the vicinity of
the first r-process abundance peak ($A \sim 80$), as those possess the
highest mass fractions. The properties of the 
isotopes which contribute most to the heating rates in the neutrino-driven
wind are listed in Tab.~\ref{tab:top-heating}. On average, about 40\% of 
the decay energy is carried away only by neutrinos. An additional fraction of energy is 
radiated away by escaping photons, while both the electrons and the remaining part of the 
photons thermalize. Therefore our hypothesis of an effective beta-decay 
thermalization fraction of 50\% is reasonable.

\begin{table}[!htb]
\begin{center}
\vspace{0.05 in}\caption{Properties of the dominant $\beta-$decay nuclei at 
$t \sim 1$ day; based on the data from NuDat 2.6 database
({\tt http://www.nndc.bnl.gov/nudat2/})}
\label{tab:top-heating}
\begin{tabular}{lccccccc}
\hline
\hline
  \multicolumn{1}{c}{Isotope} &
  \multicolumn{1}{c}{$t_{1/2}$} &
  \multicolumn{1}{c}{Q$^{(a)}$} & 
  \multicolumn{1}{c}{$\epsilon_{e}^{(b)}$} &
  \multicolumn{1}{c}{$\epsilon_{\nu}^{(c)}$} &
  \multicolumn{1}{c}{$\epsilon_{\gamma}^{(d)}$} &
  \multicolumn{1}{c}{$E_{\gamma}^{\rm avg\,(e)}$} & 
\\
 & (h) & (MeV) & & & & (MeV) \\
\hline
  $^{88}$Kr  & 2.83 & 2.92 & 0.12 & 0.21 & 0.67 & 1.34 \\
  $^{88}$Rb  & 0.30 & 5.31 & 0.39 & 0.49 & 0.13 & 1.59 \\
  $^{87}$Kr  & 1.27 & 3.89 & 0.34 & 0.46 & 0.20 & 0.95 \\
  $^{83}$Br  & 2.37 & 0.98 & 0.33 & 0.66 & 0.007& 0.46 \\
  $^{81}$Sr  & 0.37 & 3.93 & 0.28 & 0.37 & 0.35 & 0.42 \\
  $^{78}$Ge  & 1.47 & 0.96 & 0.24 & 0.47 & 0.29 & 0.28 \\
  $^{78}$As  & 1.51 & 4.21 & 0.30 & 0.39 & 0.31 & 0.94 \\
  $^{77}$Ge  & 11.2 & 2.70 & 0.23 & 0.36 & 0.41 & 0.47 \\
\hline
\end{tabular}
\end{center}
{\small
(a) Total energy released in the decay; 
(b),(c),(d) Fraction of the decay energy released in electrons, neutrinos, and $\gamma-$rays; 
(e) Average photon energy produced in the decay.}
\end{table}

\begin{figure}[!htb]
  \begin{center}
    \includegraphics[width=0.95\linewidth]{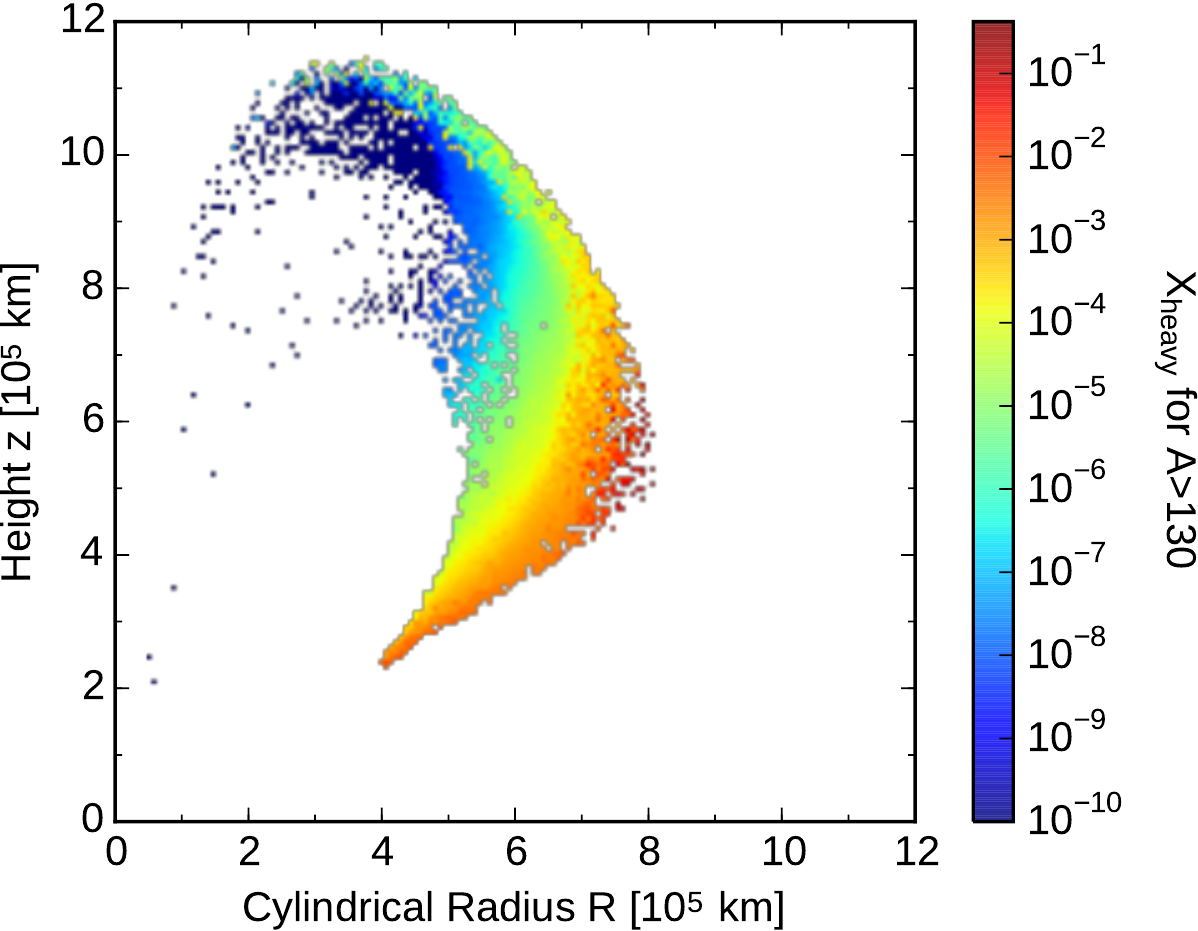}
  \end{center}
  \caption{Distribution of summed up mass fractions for nuclei with mass number $A > 130$ in the neutrino-driven wind. 
    The upper half space is shown in cylindrical coordinates for a time of 50~s after the
    merger. Tracers with vanishing total mass fractions $X_\mathrm{heavy}<10^{-10}$ are 
    set to $10^{-10}$.}
  \label{fig:mf_distribution}
\end{figure}

Here we extend the semi-analytic model of \cite{Grossman.etal:2014} and 
\cite{Perego.etal:2014} to compute light curves. We assume that the wind 
shuts off immediately after the collapse of the MNS and we explore collapse 
times of 90~ms, 140~ms and 190~ms after merger. The masses as well as average 
properties of the ejecta in the four bins for each of the cases are given 
in Tab.~\ref{tab:bin-props}. The higher value of $\kappa = 10\;{\rm cm}^2/{\rm g}$ 
for the opacity in the last angular bin is justified by 
inspecting the distribution of mass fractions for heavy elements $X_{\rm heavy}$ 
(for mass numbers $A>130$) in the neutrino-driven wind. This distribution is
portrayed for extrapolated positions of the unbound tracers in 
Fig.~\ref{fig:mf_distribution} at $t = 50~{\rm s}$ after the merger. 
Mass fractions of heavy elements are marginal in the bins~$1-3$, but 
heap up in the area of bin~4. Reflecting that sizeable fractions of heavy 
elements are present, we adopt an opacity similar to the one of 
lanthanides or actinides \citep{Kasen.etal:2013} for the last bin.


\begin{table}[!htb]
 \begin{center}
  \caption{Parameters used for computing luminosities of individual bins:
   spanned solid angle, mass of the bin, average asymptotic velocity of the
   ejecta in the bin, and the adopted value of the gray opacity.}
  \label{tab:bin-props}
  \begin{tabular}{lccccc}
  \hline
  $t_{\rm sim}$ & bin 
         & $\Delta\Omega$  
         & $\Delta m_{\rm ej}/10^{-3}M_\sun$  
         & $v_{\rm ej}/c$
         & $\kappa$ $[{\rm cm}^2/{\rm g}]$  \\
  \hline
  $90\ {\rm ms}$  
      & 1 & 0.21409 & $0.013$ & 0.05055 & 1 \\
      & 2 & 0.62769 & $0.30$  & 0.07974 & 1 \\
      & 3 & 0.99851 & $0.77$  & 0.07287 & 1 \\
      & 4 & 1.30129 & $0.34$  & 0.06808 & 10\\
  \hline 
  $140\ {\rm ms}$
      & 1 & 0.21409 & $0.05$ & 0.04655 & 1 \\
      & 2 & 0.62769 & $0.75$ & 0.07483 & 1 \\
      & 3 & 0.99851 & $1.99$ & 0.07626 & 1 \\
      & 4 & 1.30129 & $2.43$ & 0.06694 & 10\\
  \hline 
  $190\ {\rm ms}$
      & 1 & 0.21409 & $0.07$ & 0.04694 & 1 \\
      & 2 & 0.62769 & $1.18$ & 0.07325 & 1 \\
      & 3 & 0.99851 & $3.39$ & 0.07500 & 1 \\
      & 4 & 1.30129 & $4.80$ & 0.06466 & 10\\
  \hline
  \end{tabular}
 \end{center}
\end{table}

\begin{figure}[!htb]
\begin{center}
  \includegraphics[width=0.95\linewidth]{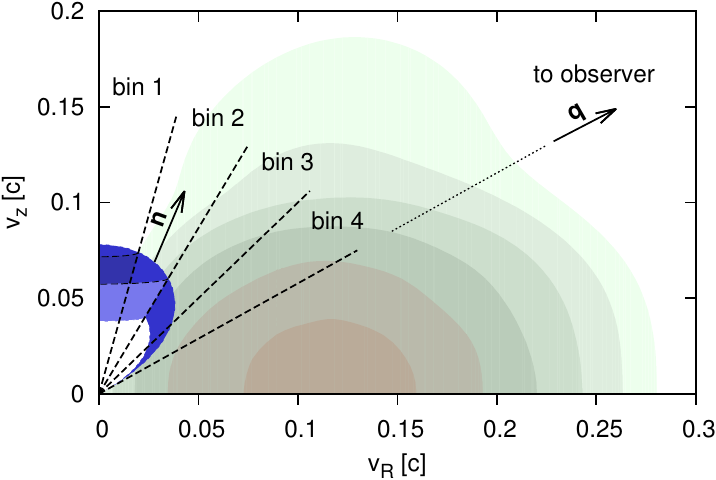} 
\end{center}
\caption{Morphology of the neutrino-driven wind (blue, this work) and dynamic ejecta (gray, cf. \cite{Rosswog.etal:2014}) in 
asymptotic velocity space, represented by a polar cut through the outflows.
Shaded areas represent the density of dynamic ejecta, spaced by 0.5 dex from
the maximum density at about $v_R\sim0.115c$. Each angular bin is approximated by a 
conical slab with unit vector $\vec n$ normal to the photosphere. An observer
is pointed to by a unit vector $\vec{q}$.}
\label{fig:model-illustration}
\end{figure}

Figure~\ref{fig:model-illustration} illustrates our semianalytic model for
light curve calculations, combining the results of this work and of 
\cite{Rosswog.etal:2014}. The neutrino-driven wind is schematically shown
in blue, while the gray-shaded areas tag the density of the dynamic ejecta 
in steps of 0.5~dex. We subdivide the wind outflow into the same four bins 
as in Section 2.1 above.
Each bin is then approximated by a conical slice of a spherically-symmetric
outflow with radial density distribution averaged over the bin and expanding
with averaged velocity (a similar approach is employed in \cite{Margalit.Piran:2015} in a context
of radio flares). All radiation is assumed to
escape from the photosphere for each bin. Vector $\vec{n}$ is the unit normal
to the photosphere, and an observer is pointed to by a unit vector $\vec{q}$.
The sketch also illustrates possible obscuration effects from the dynamic, very 
opaque ejecta, which are concentrated in a puffed up toroid around the equatorial plane.
It is apparent that when viewed from the equatorial plane, the wind
emission is completely obscured, while from the pole it is possible to observe
at least the first three bins.

For a bin $k$ with mass $\Delta m_k$ which spans solid angle $2\Delta\Omega_k$
(a factor of 2 takes into account upper and lower lobes), we take a
spherically-symmetric outflow with mass $4\pi\Delta m_k/2\Delta\Omega_k$ and
compute its isotropic bolometric luminosity $L_{k,{\rm iso}}$ as described
in~\cite{Grossman.etal:2014} (see their Sect.~4.1).

The luminosity is generated by radioactive heating in the bulk of the
outflow and the resulting photons escape from a photosphere. Spanning
the same solid angle $2\Delta\Omega_k$ for each bin, the luminosity for
bin $k$ is a proportional fraction of isotropic luminosity of the spherical model:
\begin{align*}
L_k= \frac{\Delta\Omega_k}{2\pi} L_{k,{\rm iso}}.
\end{align*}

\begin{figure}[!htb]
\begin{center}
  \includegraphics[width=0.95\linewidth]{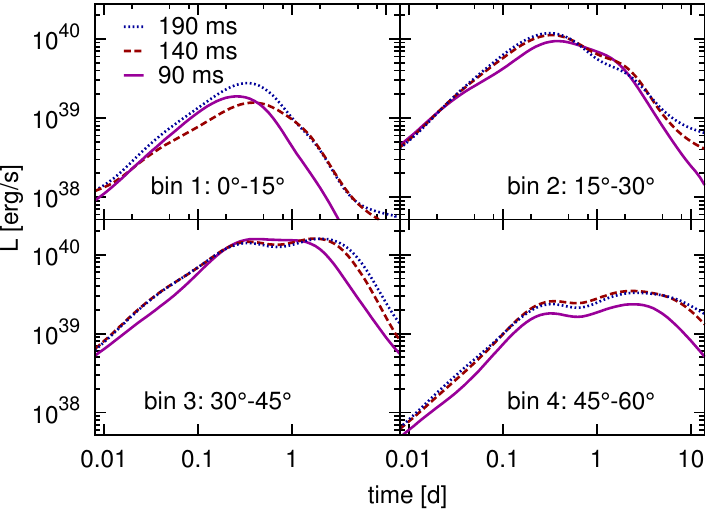} 
\end{center}
\caption{Individual contributions of the combined ejecta to the luminosity in the angular bins for three
different MNS collapse times: 90~ms (solid), 140~ms (dashed) and 190~ms
(dotted line).}
\label{fig:lightcurves-bins}
\end{figure}

We then compute the total luminosity of the wind outflow by summing up
individual bin contributions, ignoring possible radiative flux between
the bins. Individual bin contributions for the combined bolometric luminosity of
the neutrino-driven wind and dynamic ejecta for three cases of the MNS 
collapse time delay are displayed in Fig.~\ref{fig:lightcurves-bins}.
For the dynamic ejecta, we used an average case (model~A
from~\cite{Grossman.etal:2014}) with mass $1.3 \cdot 10^{-2}\;M_\sun$. 
Notice that the luminosity of bin~1 is dimmer than for the bins~2
and 3, because it spans a smaller solid angle. 
The contribution from bin~4 is not only smaller, but also peaks much later ($3-4$ days)
due to the high opacity caused by the very heavy nuclear content \citep{Kasen.etal:2013},
such that the medium becomes delayed transparent when the temperature has decreased.

\begin{figure}[!htb]
\begin{center}
  \includegraphics[width=0.95\linewidth]{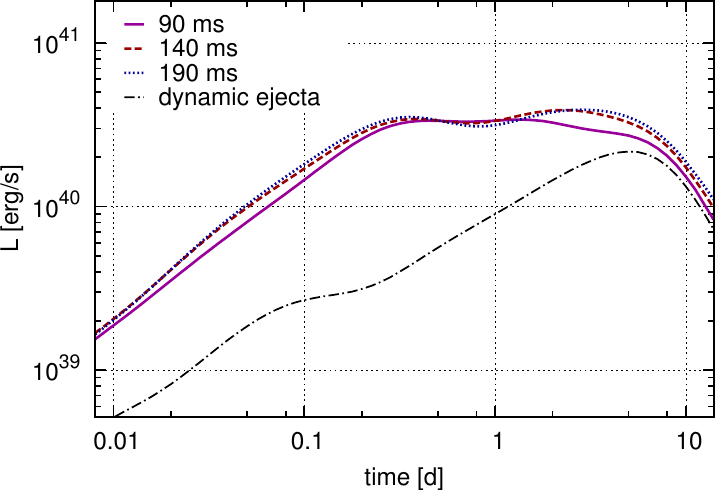} 
\end{center}
\caption{Bolometric luminosities for three cases of MNS collapse times: 90~ms
(solid), 140~ms (dashed) and 190~ms (dotted line), computed by summing up
contributions from individual bins and adding the contribution from dynamic
ejecta. Contribution from the latter is also plotted separately (dash-dotted
line).}
\label{fig:lightcurves-total}
\end{figure}

Combined bolometric luminosities of the neutrino-driven wind and dynamic ejecta are shown 
in Fig.~\ref{fig:lightcurves-total}. A first peak of $L_{\rm peak} \sim 4 \cdot 10^{40}$~erg/s is reached at 
about 4~h after the merger.
This luminosity stays roughly constant for several days while rapidly shifting into
the infrared band, as shown below. 
For higher MNS collapse times, the light curve exhibits a double peak structure,
but overall the bolometric light curves on the plots exhibit no appreciable
difference between different MNS collapse times.

\begin{figure}[!htb]
\begin{center}
  \includegraphics[width=0.95\linewidth]{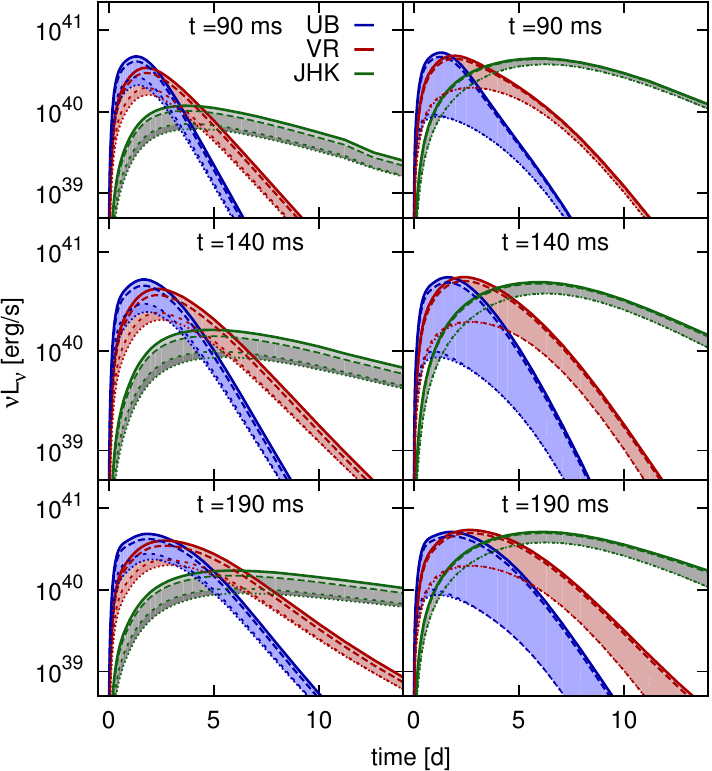} 
\end{center}
\caption{Broadband light curves of the wind outflow (left panel) and
wind+dynamic ejecta (right panel), showing the combined blue (U+V bands),
red (V+R) and infrared (J+H+K) contributions. The top, middle and bottom rows
show the three cases of MNS collapse times: 90~ms, 140~ms and 190~ms,
respectively. The range for each light curve spans possible inclination angles
of the system axis with respect to the observer: $0^\circ$ (solid) -- view
from the top, $30^\circ$ (dashed), $60^\circ$ (short dashes) and $90^\circ$
(dotted) -- view "edge-on".
}
\label{fig:broadband-lightcurves}
\end{figure}

Our simple model also allows to calculate approximate light curves in
different bands for different orientations of the system with respect to an
observer. At any given moment $t$ we can compute an effective temperature
$T_{k,{\rm eff}}$ for a bin $k$ using the formula:
\begin{align}
L_{k,{\rm iso}}(t)= 4\pi r_{k,{\rm ph}}^2 \cdot 
                    (\sigma_{\rm SB}T_{k,{\rm eff}}^4)
\end{align}
where $r_{k,{\rm ph}}(t)$ is the photospheric radius for the spherical model
of bin $k$, and $\sigma_{\rm SB}$ is the Stefan-Boltzmann constant.
We make an assumption that all the flux is emitted from the photosphere with
blackbody spectrum $B_{\nu}(T_{k,{\rm eff}})$ and isotropic intensity
(following Lambert's law).
Then, the spectral flux in the direction $\vec{q}$ is given by: 
\begin{align}
\mathcal{F}_{\nu}(\vec{q},t)= \sum_k B_{\nu}(T_{k,{\rm eff}}(t))
                      \iint_{(\vec{n}_k\cdot\vec{q})>0} 
                             (\vec{q}\cdot d\vec{\Omega}), 
\end{align}
where $\vec{n}_k$ is unit normal to the photosphere of bin $k$, and
integration for each bin spans the part of the surface facing the observer
(stated by the condition $(\vec{n}_k\cdot\vec{q})>0$). This is also
illustrated in Fig.~\ref{fig:model-illustration}.
Essentially, the integrals in this formula are simply time-independent
geometric projections of the photosphere for each bin onto the viewing plane.
They can be computed beforehand and used as weighting factors $p_{k}(\vec{q})$
for calculating the combined light curve:
\begin{align}
\mathcal{F}_{\nu}(\vec{q},t)= \sum_k p_{k}(\vec{q}) B_{\nu}(T_{k,{\rm eff}}(t)).
\end{align}
After integrating $\mathcal{F}_{\nu}(\vec{q},t)$ over certain frequency ranges, we
obtain broadband light curves, shown in Fig.~\ref{fig:broadband-lightcurves}.
The left-hand panel shows synthetic light curves for wind outflow only,
while the right-hand panel displays the combined contribution from both wind
and dynamic ejecta. 

Orientation effects are displayed as a range of values for each
band, with the maximum magnitude achieved when the system is viewed
"face-on", i.e. from the pole. As pointed out by \cite{Kasen.etal:2015},
even a small amount of dynamic ejecta can completely obscure the
optical emission from the winds. Therefore, when adding up
contributions from different bins and dynamic ejecta, we emulate
this obscuration by excluding certain bins, depending on the
inclination angle $\theta$ with respect to the observer. Our choice
of which bins to exclude is motivated by the geometry of the ejecta
(depicted schematically in Fig.~\ref{fig:model-illustration}). Specifically, we assume that 
bin~4 is completely obscured for all observing angles; bins~2 and 3 are
obscured when $\theta\ge60^{\circ}$, and for the edge-on view with
$\theta=90^{\circ}$ the wind outflow emission is obscured completely.
For this reason, depending on system orientation, the luminosity in
different bands can vary by up to an order of magnitude, from bright
blue when observed from the pole, to dim infrared when observed from
the side.

\section{Discussion and conclusions}
\label{sec:conclusions}

We have presented a comprehensive study of the neutrino-driven
wind nucleosynthesis in the aftermath of a double neutron star
merger. We have focused on the initial phase where the remnant
consists of a MNS surrounded by a massive accretion disk. The
current nucleosynthesis study is based on the first three-dimensional
simulation of such a wind \citep{Perego.etal:2014}, sampling the wind outflow
by $\sim 17\,000$ tracer particles. For every tracer, nucleosynthesis
calculations are performed in a post-processing step. 
This work advances the preliminary study in \cite{Perego.etal:2014} that
only used 10 tracers and thus led to a slight overestimation
of the amount of heavy elements ($A>130$) produced, whereas the 
total mass ejected in the wind was underestimated.
Our main results are summarized and discussed alongside their uncertainties in the following.

The \emph{abundance distribution} in the neutrino-driven wind is
characterized by the production of lighter heavy elements
($A\lesssim 130$). We have found a time and angle dependency of the
composition correlated with the electron fraction
evolution. Elements heavier than $A>130$ are produced only at early
times, in an almost negligible amount and in the wind region that is closest
to the disk (between $45^\circ$ and $60^\circ$ from the pole, bin 4
in Fig.~\ref{fig:integrated-nucleosynthesis-140ms}). In contrast, matter ejected more perpendicular
to the disk (bin 1: $0^\circ - 15^\circ$) produces only nuclei
up to the first r-process peak ($A<100$). 
The nucleosynthesis path of the neutrino-driven wind after a
neutron star merger does not reach the heaviest nuclei, in some
cases it even stops before $N=82$. Relevant nuclear physics
input (e.g., nuclear masses and beta-decays) along the path is known
or will be measured in current and future generation facilities in
the coming years to provide more reliable data. Our calculations are based on the FRDM mass model
\citep{Moeller.etal:1995} and consistently calculated nuclear reaction rates
\citep{Rauscher.Thielemann:2000}. A detailed study of the impact of the
nuclear physics input can give rise to new insights \citep{Mendoza-Temis.etal:2014,Eichler.etal:2015,Goriely.etal:2015}. 
Especially near the shell closure at $N=82$ nuclear masses and neutron
captures may have an impact on the abundances before the second
r-process peak where we find a significant trough.

The \emph{electron fraction} is here the key parameter to understand
the characteristics of abundances. It covers a broad range from $0.1$ to
$0.4$ with average values of $Y_{\rm e}\sim 0.33$. As noted already in earlier
studies \citep[see in particular Fig. 8 in \cite{Korobkin.etal:2012} where the
$Y_{\rm e}$-dependence was explored;][]{Freiburghaus.etal:1999b,Kasen.etal:2015}, 
$Y_{\rm e} \approx 0.25$ is the threshold beyond which no more heavy r-process elements are
formed. Since abundances depend strongly on $Y_{\rm e}$, an accurate
treatment of weak reactions is crucial 
and can be achieved by improving the neutrino transport and including more
neutrino-matter reactions. We have estimated a $\sim$~20\% error on the
electron fraction due to the approximate transport scheme. This is suggested by
the comparison of neutrino luminosities and mean energies from
\cite{Perego.etal:2014} and \cite{Dessart.etal:2009}, which are among the best treatments for these
kinds of systems (see also \cite{Just.etal:2015a, Just.etal:2015b, Fernandez.Metzger:2013}). For the neutrino-matter
interactions, a small variation of $Y_{\rm e}$ can arise from a detailed
inclusion of the (anti)neutrino absorption including weak
magnetism and medium effects \citep{Roberts.etal:2012, Martinez-Pinedo.etal:2012}.

The \emph{amount of ejecta} also depends on angle and time. As the 
wind is preferentially driven out at lower latitudes, bin 3 and 4 contain
roughly 6 times the matter of bins 1 and 2. About 110~ms after the merger
a steady-state is reached in terms of mass loss.
The total unbound mass depends crucially on the time when the MNS
collapses and if it survives for $\sim 200$~ms as much as $9 \cdot
10^{-3}M_\odot$ can become unbound (i.e. $\sim 5$\% of the initial 
mass of the accretion disk). This amount of matter rivals the
dynamic ejecta that are typically found to be of order of $0.01 M_\odot$
($1.3 \cdot 10^{-2}M_\odot$ for the specific case considered here).
Recent studies \citep{Fernandez.Metzger:2013,Kasen.etal:2015,Just.etal:2015a,Sekiguchi.etal:2015}
indicate that the winds switches off rapidly once a black hole forms.
Approximately half of the
neutrino luminosity is provided by the MNS. Therefore, when the
black hole forms this contribution vanishes and the disk structure may
significantly change. It is possible that general relativistic
effects also have a non-negligible impact on the disk winds from a
MNS and accretion disk system, as is the case for neutron star -
black hole systems \citep{Caballero.etal:2015}.

The angle dependency of the composition may have consequences for
the \emph{mixing} with other ejecta. Both dynamic and viscous ejecta
contribute also to the production of heavy r-process elements beyond
the second peak ($A>130$). If wind material perfectly mixes with
these ejection channels, its contribution may lead to variations for the lighter heavy
elements (i.e., $A<130$). This variation is indeed observed in stars
with high enrichment of heavy r-process elements. However, if the
mixing is not perfect one could speculate that neutron star mergers
with long-lived MNS may produce different abundance patterns
including one with low enrichment of heavy r-process elements. This
kind of pattern (``Honda-like pattern'', see \cite{Honda.etal:2004,Honda.etal:2007}) 
is observed in a few very old
stars and its origin is still unknown. However, detailed mixing
models following the late evolution of the three
nucleosynthesis-relevant ejecta are required before concluding about
the contribution of neutron star mergers to observed abundance
patterns.

The initial separation of wind and dynamic ejecta allows to describe 
the individual contributions to the \emph{light curve} of the 
electromagnetic transient. We use a semi-analytic radiation transport 
model to calculate the luminosities of the wind and dynamic ejecta. The 
light curve of the wind ejecta peaks after $\sim 4$~hours in the blue 
compared to the one of the dynamic ejecta that peaks few days after the merger in the 
infrared. 
The combined light curve will significantly depend on the observing
angle, as previously pointed out by \cite{Kasen.etal:2015}. Due to the high
line expansion opacities of lanthanides and actinides
\citep{Kasen.etal:2013}, even small amounts of dynamic ejecta along the
line of sight can completely obscure the blue transient from the
wind. Since the morphology of dynamic ejecta is such that it leaves
polar regions evacuated, it is far more likely to detect the blue
component when the system is observed from the pole. In Fig.~\ref{fig:broadband-lightcurves}, we
have presented the light curves of obscured and unobscured winds as a
function of angle and time of black hole formation. The luminosity in the
blue band exhibits a much higher dependency on the viewing angle
compared to the luminosity in the infrared, in agreement with previous
studies \citep{Kasen.etal:2015}. This significant anisotropy needs to be
taken into account when assessing detectability of optical
counterparts to neutron star mergers. Another interesting feature is the noticeable dependence of the
position of optical peaks on the time of black hole formation
(apparent in Fig.~\ref{fig:broadband-lightcurves}). A longer collapse time leads to a later and more
prolonged peak in both U+B and V+R bands, while this trend is washed
out in the infrared. Although at this point it is difficult to draw any
quantitative conclusions due to limitations of our model, our results
suggest a possibility to use the position and duration of optical
transients as proxies for the time of black hole formation (see also
\cite{Metzger.Fernandez:2014}). More work is needed to take into account
additional channels of late-time ejecta production, and to improve
radiation treatment (see, e.g., \cite{Kasen.etal:2015}).
\newline

Our results clearly indicate the importance of neutrino-driven winds for
the nucleosynthesis of neutron star mergers. It needs to be stated, however, that these results
have been obtained based on Newtonian hydrodynamics simulations
and that general-relativistic effects could potentially modify some
of our conclusions. For example, recent studies in full GR \citep{Hotokezaka.etal:2011,Hotokezaka.etal:2013,Sekiguchi.etal:2015} and in the
conformal flatness approximation \citep{Bauswein.etal:2013} agree that shocks
are more important for the dynamic ejecta in the GR case, especially when a soft equation of
state is used. This is mainly because the stars are -- for a given EOS --
more compact due to the stronger relativistic gravity and hence
lead to larger velocities prior to contact. A softer EOS leads compared
to a stiff one to more compact stars and also to smaller sound speeds
which, in turn, make the occurrence of shocks more likely. Therefore,
the dynamic ejecta are dominated by the hotter "interaction component"
while Newtonian ejecta are dominated by the colder, unshocked "tidal
component" \citep{Korobkin.etal:2012} \footnote{Despite these differences,
the overall amount of dynamic ejecta is in good agreement between
different studies. There is a dependence on the mass ratio and the
equation of state, with softer EOSs ejecting more mass in GR calculations,
but ejecta masses up to a few percent of the solar mass can be reached, see, e.g. Tab. 1 in
\cite{Rosswog:2015} for a compilation of the ejecta mass results from
different groups. It can, however, not be excluded that also purely
numerical effects still impact on the results. 
}. Therefore, a GR calculation will plausibly cause a higher
temperature environment which is expected to lead to larger electron fractions
in the ejecta. Due to shock-heating it is also likelier that unbound material 
is ejected out-of-plane. Thus a neutron star merger could actually, consistent
with our findings here, produce a very broad range of r-process abundances
\citep{Wanajo.etal:2014,Just.etal:2015a}.

Understanding the interplay of all ejecta from neutron star mergers is key to decode
the diverse aspects of this astrophysical site. Further detailed simulations and
complete nucleosynthesis studies will help to pin down the role of neutron star mergers for 
the origin of the heavy elements in the universe as well as to predict reliable
electromagnetic counterparts for the detection of GWs.


\begin{acknowledgements}
AA, AP, and DM are supported by
 the Helmholtz-University Young Investigator grant
 No. VH-NG-825. AP acknowledges the use of computational resources provided by the Swiss Super Computing
Center (CSCS), under the allocation grant s414.
FKT is supported by the Schweizerische Nationalfond (SNF) and the ERC Advanced Grant FISH.
OK and SR were supported by Deutsche Forschungsgemeinschaft (DFG) 
 under grant number RO-3399/5-1 and by the Swedish Research Council (VR) 
 under grant 621-2012-4870. Some of the simulations have been obtained
 on the facilities of the The North-German Supercomputing Alliance (HLRN).
The authors thank also "NewCompStar", COST Action MP1304, for support.
We also thank the Institute for Nuclear Theory where discussions and 
contributions to this work were made during the program "INT-14-2b" 
about nucleosynthesis and chemical evolution.
\end{acknowledgements}

\nocite{*}
\bibliographystyle{apj}
\bibliography{bibliography}

\end{document}